\documentclass[reprint,amsmath,amssymb,aps]{revtex4-1}
\usepackage[latin9]{inputenc}
\usepackage{color}
\usepackage{units}
\usepackage{amsmath}
\usepackage{graphicx}

\makeatletter

\providecommand{\tabularnewline}{\\}

\usepackage{caption}
\usepackage{dcolumn}
\usepackage{bm}
\usepackage{units}

\@ifundefined{showcaptionsetup}{}{%
 \PassOptionsToPackage{caption=false}{subfig}}
\usepackage{subfig}
\makeatother

\begin{document}
~

\title{Maximum Entropy Principle Analysis in Network Systems with Short-time
Recordings}

\author{Zhi-Qin John Xu$^{1}$}

\author{Jennifer Crodelle$^{2}$}

\author{Douglas Zhou$^{3}$}

\thanks{zdz@sjtu.edu.cn}

\author{David Cai$^{1,2,3,4}$}

\affiliation{$^{1}$NYUAD Institute, New York University Abu Dhabi, Abu Dhabi,
United Arab Emirates,~\\
 $^{2}$Courant Institute of Mathematical Sciences, New York University,
New York, New York, USA.~\\
 $^{3}$School of Mathematical Sciences, MOE-LSC and Institute of
Natural Sciences,Shanghai Jiao Tong University, Shanghai, P.R. China,~\\
 $^{4}$Center for Neural Science, New York University, New York,
New York, USA.~\\
}

\date{\today}
\begin{abstract}
In many realistic systems, maximum entropy principle (MEP) analysis
provides an effective characterization of the probability distribution
of network states. However, to implement the MEP analysis, a sufficiently
long-time data recording in general is often required, e.g., hours
of spiking recordings of neurons in neuronal networks. The issue of
whether the MEP analysis can be successfully applied to network systems
with data from short recordings has yet to be fully addressed. In
this work, we investigate relationships underlying the probability
distributions, moments, and effective interactions in the MEP analysis
and then show that, with short recordings of network dynamics, the
MEP analysis can be applied to reconstructing probability distributions
of network states under the condition of asynchronous activity of
nodes in the network. Using spike trains obtained from both Hodgkin-Huxley
neuronal networks and electrophysiological experiments, we verify
our results and demonstrate that MEP analysis provides a tool to investigate
the neuronal population coding properties, even for short recordings. 
\begin{description}
\item [{PACS numbers}] 89.70.Cf, 87.19.lo, 87.19.ls, 87.19.ll 
\end{description}
\end{abstract}

\pacs{Valid PACS appear here}
\maketitle

\section*{Introduction}

Binary-state models have been used to describe the activity of nodes
in many network systems, such as neuronal networks in neuroscience
\cite{schneidman2006weak,shlens2006structure,xu2016dynamical}. Understanding
the distribution of network binary-state dynamics is important in
unveiling underlying network function, especially in neuroscience
where both theoretical and experimental results indicate that populations
of neurons perform computations probabilistically through their firing
patterns \cite{knill2004bayesian,karlsson2009awake}. For instance,
statistical distributions of neuronal network firing patterns have
been shown to perform awake replays of remote experiences in rat hippocampus
\cite{karlsson2009awake}. Therefore, studying the characteristics
of neuronal firing pattern distributions will help to understand how
neuronal networks encode information \cite{morcos2016history}. However,
this is a difficult task, since the number of all possible network
states grows exponentially as the network size increases, i.e., $2^{n}$
for a network of $n$ binary-state nodes. This high dimensionality
presents a challenge in directly measuring the distribution of network
states in electrophysiological experiments, especially for the case
of \emph{in vivo} measurements on awake animals, thus, the difficulty
in understanding coding schemes in neuronal networks.

The maximum entropy principle (MEP) analysis is a statistical method
used to infer the least biased probability distribution of network
states by maximizing the Shannon entropy with given constraints, i.e.,
moments up to certain order \cite{jaynes1957information}. The inferred
probability distribution is called the MEP distribution. For example,
under constraints of low-order moments, the MEP distribution gives
rise to an accurate estimate of the statistical distribution of network
states in many scientific fields, e.g., neuroscience \cite{schneidman2006weak,shlens2006structure,tang2008maximum,xu2016dynamical},
biology \cite{sakakibara2009protein,sulkowska2012genomics,mantsyzov2014maximum},
imaging science \cite{saremi2013hierarchical}, economics \cite{golan1996maximum,squartini2013early},
linguistics \cite{stephens2010statistical}, anthropology \cite{hernando2013workings},
and atmosphere-ocean science \cite{khatiwala2009reconstruction}.

However, a practical and important issue remains unclear: how well
the MEP analysis performs when the recording time of dynamics of network
nodes is short. This is because a very long recording is often required
to carry out the MEP analysis, e.g., hours for a network of $10$
neurons \cite{schneidman2006weak}. These long recordings are usually
impractical to achieve due to cost or capability. For example, physiological
constraints such as fatigue of neurons makes it very difficult to
record the state of neuronal networks over a long time, especially
\emph{in vivo} recordings on awake animals. Meanwhile, data obtained
by short recordings poorly captures many activity states, leading
to incorrect descriptions for the probability distribution of network
states. The insufficient measurements due to short recordings could
lead to a misunderstanding of information coding structure embedded
in network activity states \cite{ohiorhenuan2010sparse}. Therefore,
it is important to estimate an accurate probability distribution of
network states from short recordings where many network activities
are underrepresented.

In this work, we demonstrate that the MEP analysis can give rise to
an accurate estimate of the probability distribution of network states
from a short-time data recording if the activity of nodes in the network
is asynchronous (asynchronous network). To achieve this, we first
show the existence of a one-to-one mapping (denoted as the full-rank
matrix ${\bf U}_{{\rm PM}}^{(n)}$ for a network of size $n$) between
the probability distribution of network states (denoted by a vector
of size $2^{n}$, $\mathbf{P}^{(n)}$) and the corresponding all moments
(denoted by a vector of size $2^{n}$, $\mathbf{M}^{(n)}$). Since
the \emph{full-order MEP distribution} (the distribution obtained
in the MEP analysis with constraints of all moments $\mathbf{M}^{(n)}$)
and the probability distribution $\mathbf{P}^{(n)}$ are the same
(they have all the same moments $\mathbf{M}^{(n)}$), we derive another
one-to-one mapping (denoted as the full-rank matrix ${\bf L}_{{\rm JP}}^{(n)}$
for a network of size $n$) between all effective interactions (Lagrange
multipliers corresponding to constraints of all moments in the expression
of the full-order MEP distribution) and the probability distribution
$\mathbf{P}^{(n)}$. These mappings show that all moments and all
effective interactions can equivalently represent a probability distribution
of network states for a general network of any size.

Next, we use the above equivalent representations to show that, in
an asynchronous network, low-order MEP analysis gives rise to an accurate
estimate of the probability distribution of network states from a
short-time recording. In an asynchronous network, the probability
of many nodes being active in one sampling time window (referred to
as \emph{highly-active states}) is very small. Through the mapping
from effective interactions to the probability distribution, ${\bf L}_{{\rm JP}}^{(n)}$,
we observe that the probability of a highly-active state can be written
as a summation of effective interactions corresponding to the constraints
of moments of those active nodes. In an asynchronous network, high-order
effective interactions (Lagrange multipliers corresponding to the
constraints of high-order moments) are usually small \cite{shlens2006structure,schneidman2006weak,tang2008maximum,cavagna2014dynamical,watanabe2013pairwise},
as compared to the low-order effective interactions (Lagrange multipliers
corresponding to the constraints of low-order moments). Then, the
probabilities of highly-active states can be well estimated by the
dominating low-order effective interactions. The MEP analysis shows
that low-order effective interactions can be estimated by low-order
moments through a widely-used iterative scaling algorithm (see Appendix
\ref{subsec:A2:-The-iterative} for details). Therefore, the low-order
moments may possibly be measured accurately using short recording
(as discussed below), and may be used to perform the MEP analysis
to obtain a good estimate of the probability distribution of network
states.

To obtain an accurate estimation of the low-order moments, we use
properties of the mapping between the probability distribution of
network states and the corresponding moments of the network, ${\bf U}_{{\rm PM}}^{(n)}$.
This mapping shows that low-order moments can be written as a summation
of all probability states in which those nodes corresponding to the
constraints of moments are active, e.g., the first-order moment of
node $i$ is a summation of all network-state probabilities in which
node $i$ is active. This summation includes both highly-active state
probabilities and low-active state probabilities in which few nodes
are active in one sampling time window. Since the probability of observing
a highly-active state is small in an asynchronous network, the low-active
state probabilities dominate the summation. This results in good estimation
for the low-order moments because low-active state probabilities can
still be well-measured in a short recording and the noise for the
estimation of low-order moments can be further reduced by the summation
of low-active state probabilities \cite{saulis2012limit}.

Although the procedure described in this work can be applied to any
asynchronous network with binary dynamics, here we use spike trains
obtained from both Hodgkin-Huxley (HH) neuronal network dynamics simulations
and electrophysiological experiments to demonstrate that one can perform
the MEP analysis to accurately estimate the probability distribution
of network states from short-time recordings. For the case of simulation
data from HH neuronal networks, we evolve the HH neuronal network
for a short run time of $\unit[120]{s}$. With constraints of the
first-order and the second-order moments of the data from this \emph{short}
recording, we obtain the distribution of network states from the MEP
analysis. To verify the accuracy of the MEP distribution, we evolve
the HH neuronal network for a long run time of $\unit[1.2\times10^{5}]{s}$
and directly measure the probability distribution of network states
for comparison. For the case of experimental data from electrophysiological
measurements, we use data recorded from primary visual cortex (V1)
in \emph{anesthetized} macaque monkeys (See Appendix \ref{sec:pvc8Exp}
for details), in which we use the data of $\unit[3824]{s}$ as a long
recording data to obtain the probability distribution of network states
and perform the second-order MEP analysis on the short-recording data
of $\unit[191.2]{s}$ ($5\%$ of the total recording). Our results
show that in both numerical simulation data and electrophysiological
experimental data, these two distributions, i.e., the second-order
MEP distribution and the distribution measured in the long recording
data, indeed agree very well with each other, whereas the probability
distribution measured from the short recording deviates significantly
from them and cannot capture many neuronal activity states. 

\section*{Methods: The MEP analysis \label{sec:The-MaxEnt-model}}

In this section, we will introduce the MEP analysis, which has been
applied to estimating the probability distribution of network states
for many network systems \cite{schneidman2006weak,shlens2006structure,tang2008maximum,bury2012statistical,cavagna2014dynamical}.
The process of performing the MEP analysis is as follows. Let $\sigma\in\{0,1\}$
denotes the state of a node in a sampling time bin, where 1 refers
to an active state and 0 an inactive state. Then, the state of a network
of $n$ nodes in a sampling time bin is denoted as $\Omega=(\sigma_{1},\sigma_{2},\cdots,\sigma_{n})\in\{0,1\}^{n}$.
In principle, to obtain the probability distribution of $\Omega$,
one has to measure all possible states of $\Omega$, i.e., $2^{n}$
states in total. For the case of a network of $n$ neurons, $\sigma$
often represents whether a single neuron fires in a sampling time
bin in the network, where $1$ corresponds to a neuron that is firing
and $0$ corresponds to a neuron in the silent state. We choose a
typical bin size of $\unit[10]{ms}$ for the MEP analysis \cite{shlens2006structure,tang2008maximum}.
The state, $\Omega$, in the neuronal network would represent the
firing pattern of all neurons in the network.

The first-order moment of the state of node $i$, $\sigma_{i}$, is
given by 
\begin{equation}
\left\langle \sigma_{i}\right\rangle =\sum_{\Omega}P(\Omega)\sigma_{i}(\text{\ensuremath{\Omega)}},\label{eq:1moment}
\end{equation}
where $P(\Omega)$ is the probability of observing the firing pattern
$\Omega$ in the recording, and $\sigma_{i}(\text{\ensuremath{\Omega)}}$
denotes the state of the $i$th node in the firing pattern $\Omega$.
The second-order moment of the state of node $i$, $\sigma_{i}$,
and node $j$, $\sigma_{j}$, is given by 
\begin{equation}
\left\langle \sigma_{i}\sigma_{j}\right\rangle =\sum_{\Omega}P(\Omega)\sigma_{i}(\text{\ensuremath{\Omega)}}\sigma_{j}(\text{\ensuremath{\Omega)}, }\label{eq:2moment}
\end{equation}
with higher-order moments obtained similarly. Note that, in Eqs. (\ref{eq:1moment},
\ref{eq:2moment}), $P(\Omega)$ can be the true probability distribution
of $\Omega$, in which case, one computes the true moments; or $P(\Omega)$
can be an \emph{observed} distribution of a finite-time recording,
in which case, one then estimates the moments.

The Shannon entropy of a probability distribution $P_{{\rm any}}(\Omega)$
is defined as 
\begin{equation}
S=-\sum_{\Omega}P_{{\rm any}}(\Omega)\log P_{{\rm any}}(\Omega).\label{eq:Entropy}
\end{equation}
By maximizing this entropy, $S$, subject to all moments up to the
$m$th-order ($m\leq n$), one obtains the \emph{$m$th-order MEP
distribution} for a network of $n$ nodes \cite{schneidman2006weak,shlens2006structure,tang2008maximum}.
Note that the $k$th-order moments consist of all the expectations
of the product of any $k$ nodes' states (e.g., the second-order moments
by Eq. (\ref{eq:2moment}) above for any pair of $i$ and $j$ with
$i\neq j$) and thus when considering the constraint of the $k$th-order
moments, there are $C_{n}^{k}$ (the number of combinations of $k$
nodes from $n$ possible choices) number of constraints of $k$th-order
moments being considered. Finally, the $m$th-order probability distribution
is obtained from the following equation, 
\begin{equation}
P_{m}(\Omega)=\frac{1}{Z}\exp\left(\sum_{k=1}^{m}\sum_{i_{1}<\cdots<i_{k}}^{n}J_{i_{1}\cdots i_{k}}\sigma_{i_{1}}\cdots\sigma_{i_{k}}\right),\label{eq:PV}
\end{equation}
where, following the terminology of statistical physics, $J_{i_{1}\cdots i_{k}}$
is called the $k$th-order effective interaction ($2\leq k\leq m$),
i.e., the Lagrange multiplier corresponding to the constraint of the
$k$th-order moment $\left\langle \sigma_{i_{1}}\cdots\sigma_{i_{k}}\right\rangle $,
and the partition function, $Z$, is a normalization factor. Eq. (\ref{eq:PV})
is referred to as the \emph{$m$th-order MEP distribution}. In practice,
one can utilize an iterative scaling algorithm (see Appendix \ref{subsec:A2:-The-iterative}
for details) to numerically solve the above MEP optimization problem
to obtain the effective interactions and thus the probability distribution
in Eq. (\ref{eq:PV}). Here, for a network of size $n$, $P_{n}(\Omega)$
is referred to as the \emph{full-order MEP distribution} subject to
moments of all orders.

\section*{Results}

The results are organized as follows. We begin by demonstrating that
there is a wide range of dynamical regimes where the neuronal network
dynamics is asynchronous. Then, we show that the probability distribution
of network states measured from a short-time recording (data recorded
from the HH network dynamics with short simulation time) cannot capture
many network activity states, and thus differs from the probability
distribution of network states measured from a long-time recording.
Note that we have verified that the HH network dynamics in a long
simulation time of $\unit[1.2\times10^{5}]{s}$ reaches the steady
state; therefore, the probability distribution of network states measured
from this long recording can well represent the true probability distribution
of network states. 

We next show that there exists a one-to-one mapping between the probability
distribution of network states and the corresponding moments of the
network. Then, we combine this mapping with the full-order MEP distribution
to show that there exists a one-to-one mapping between all effective
interactions and the probability distribution of network states. Using
these mappings, we further demonstrate that high-order effective interactions
are small in asynchronous networks; thus, to accurately estimate the
probability distribution of network states, one may only require accurate
estimation of the low-order effective interactions. Finally, we make
use of low-order moments measured in short-time recordings to estimate
low-order effective interactions and show that the obtained probability
distribution from the low-order MEP analysis agrees well with the
probability distribution of network states. This is demonstrated by
both numerical simulations of asynchronous HH neuronal network dynamics
and also electrophysiological experiments.

\subsection*{Short-time recordings cannot represent all network states}

In this section, we first use numerical simulation data from the HH
neuronal network dynamics and show that short-time recordings are
often insufficient to accurately estimate the probability distribution
of network states. Later, we will also demonstrate this issue using
experimental data from electrophysiological measurements. 

We simulate a network of $80$ excitatory and $20$ inhibitory HH
neurons, with a $20\%$ connection density among neurons in the network.
As physiological experiments can often only measure a subset of neurons,
we randomly select $10\%$ of the neurons in the network ($10$ neurons)
and demonstrate differences in the directly measured probability distribution
between the short and the long recording. Note that there are typically
three dynamical regimes for neuronal networks \cite{zhou2014granger}:
(i) a highly fluctuating regime where the input rate, $\mu$, is low
(Fig. \ref{fig:Raster}a); (ii) an intermediate regime where $\mu$
is moderately high (Fig. \ref{fig:Raster}b); (iii) a low fluctuating
or mean-driven regime where $\mu$ is very high (Fig. \ref{fig:Raster}c).
We evolve the HH neuronal network dynamics and record the spike trains
of all neurons for a duration of $\unit[1.2\times10^{5}]{s}$, which
is sufficiently long to obtain a stable probability distribution of
neuronal firing patterns. We then compare the probability distribution
of network states directly measured in the short recording of $\unit[120]{s}$
to that measured in the long recording. As shown in Fig. \ref{fig:short},
the measured probability distribution of firing patterns in the short
recording deviates substantially from that in the long recording,
for all three dynamical regimes.

The probability distribution of network states is important for a
complete understanding of the underlying function of networks \cite{knill2004bayesian,karlsson2009awake}
despite the fact that a sufficient long-time data recording is often
impossible or impractical. Thus, it is essential to obtain an accurate
estimate of the probability distribution of network states from a
short-time recording. To achieve this, we next study relationships
among the probability distribution, the moments, and the effective
interactions in the MEP distribution and show that, for asynchronous
networks, the second-order MEP analysis gives rise to an accurate
estimate of the probability distribution of network states.
\begin{center}
\begin{figure*}
\subfloat[]{\begin{centering}
\includegraphics[scale=0.37]{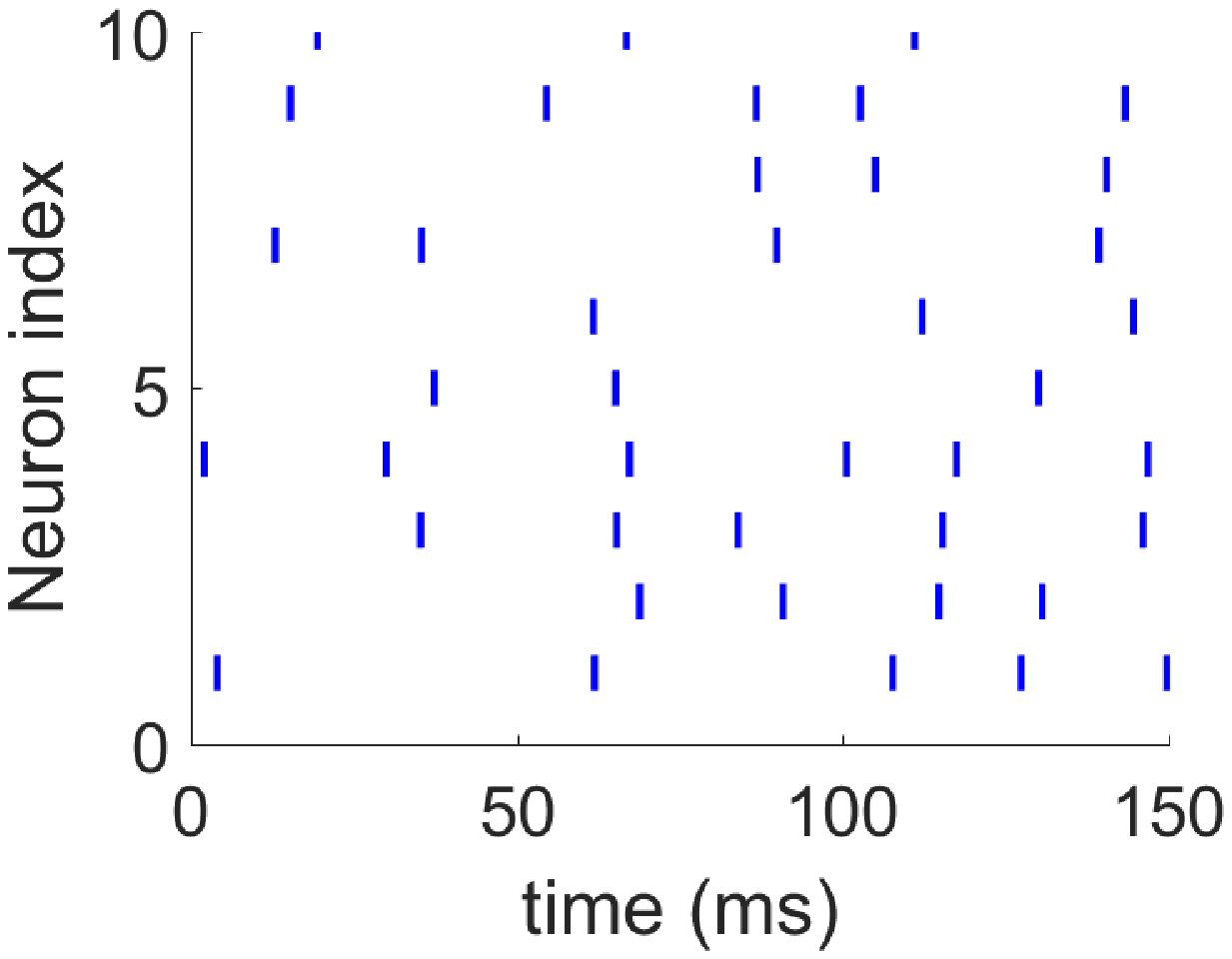} 
\par\end{centering}
}\subfloat[]{\begin{centering}
\includegraphics[scale=0.37]{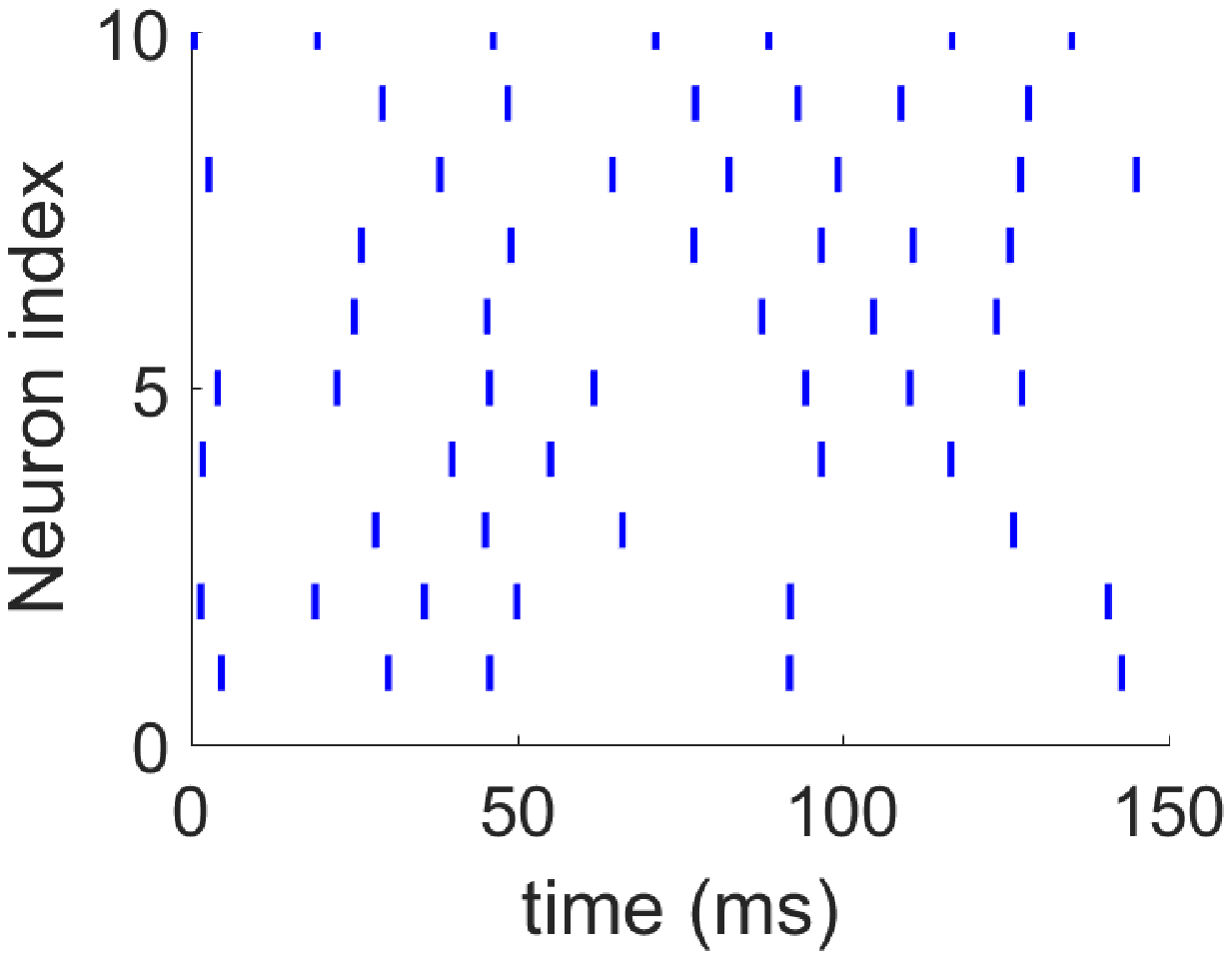} 
\par\end{centering}
}\subfloat[]{\begin{centering}
\includegraphics[scale=0.37]{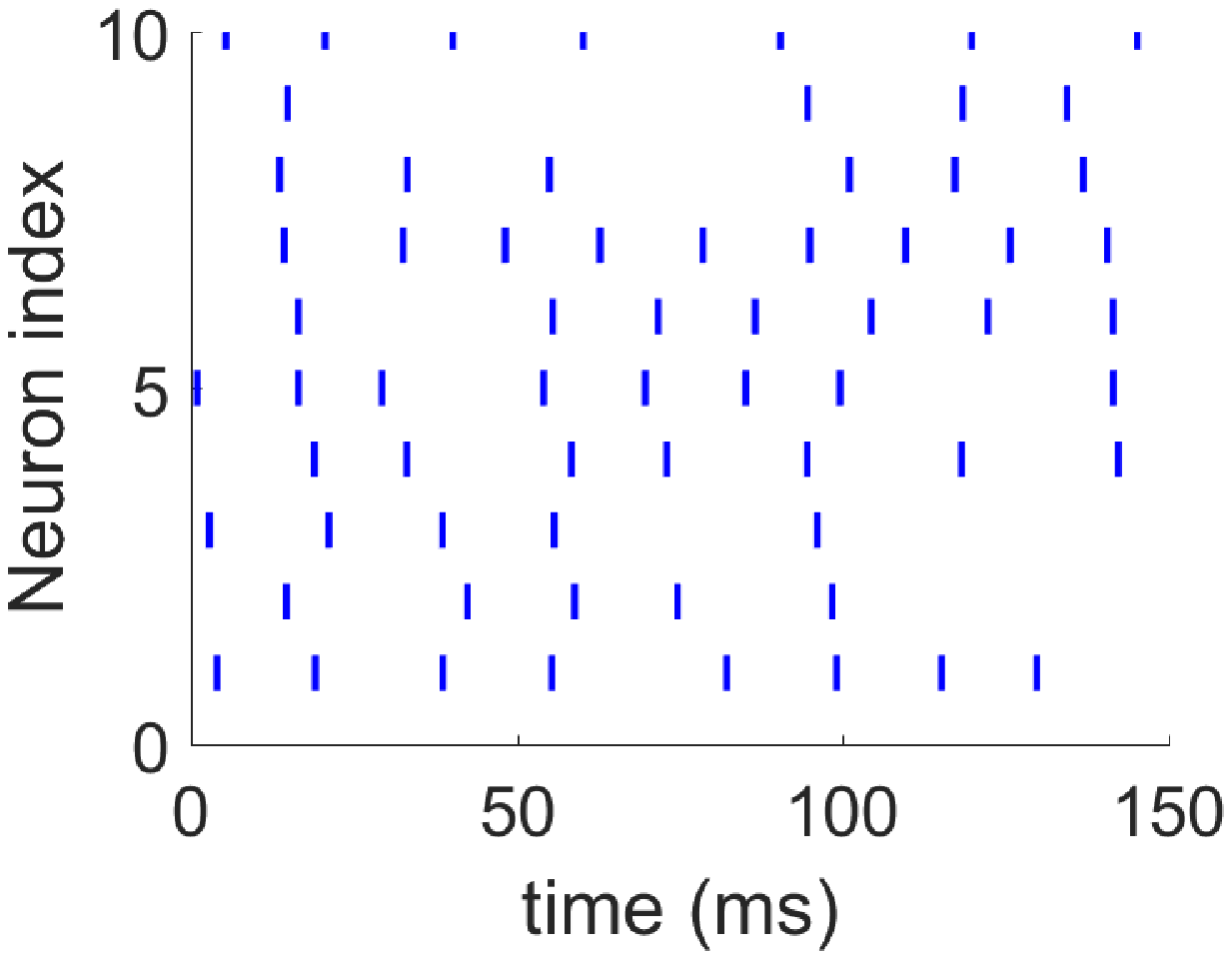} 
\par\end{centering}
}

\caption{\textbf{Raster plots for the HH neuronal network in three different
dynamical regimes.} Raster plots of $10$ randomly selected neurons
for each case are shown. A short bar indicates that the neuron with\textcolor{red}{{}
}certain index fires at certain time. The coupling strength is selected
at random from the uniform distribution of the interval $[0,s]$,
where $s=\unit[0.071]{ms^{-1}}$ (the corresponding physiological
excitatory postsynaptic potential is $\sim\unit[1]{mV}$). The Poisson
input parameters for in (a), (b), and (c) are $(\mu=\unit[0.6]{ms^{-1}},\,f=\unit[0.05]{ms^{-1}})$,
$(\mu=\unit[1.1]{ms^{-1}},\,f=\unit[0.04]{ms^{-1}})$ and $(\mu=\unit[2.5]{ms^{-1}},\,f=\unit[0.03]{ms^{-1}})$,
respectively. \label{fig:Raster}}
\end{figure*}
\par\end{center}

\begin{center}
\begin{figure*}
\subfloat[]{\begin{centering}
\includegraphics[scale=0.42]{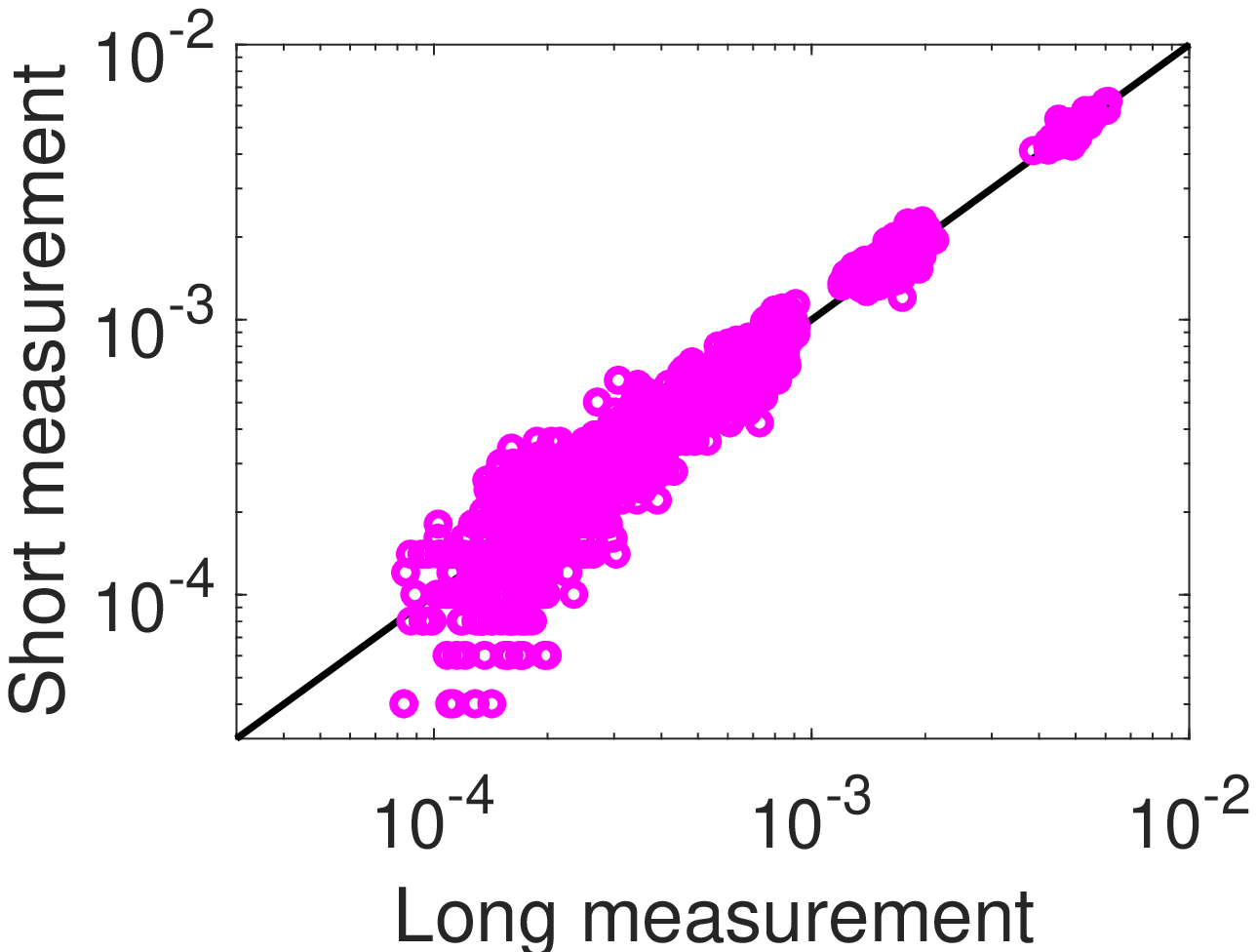} 
\par\end{centering}
}\subfloat[]{\begin{centering}
\includegraphics[scale=0.42]{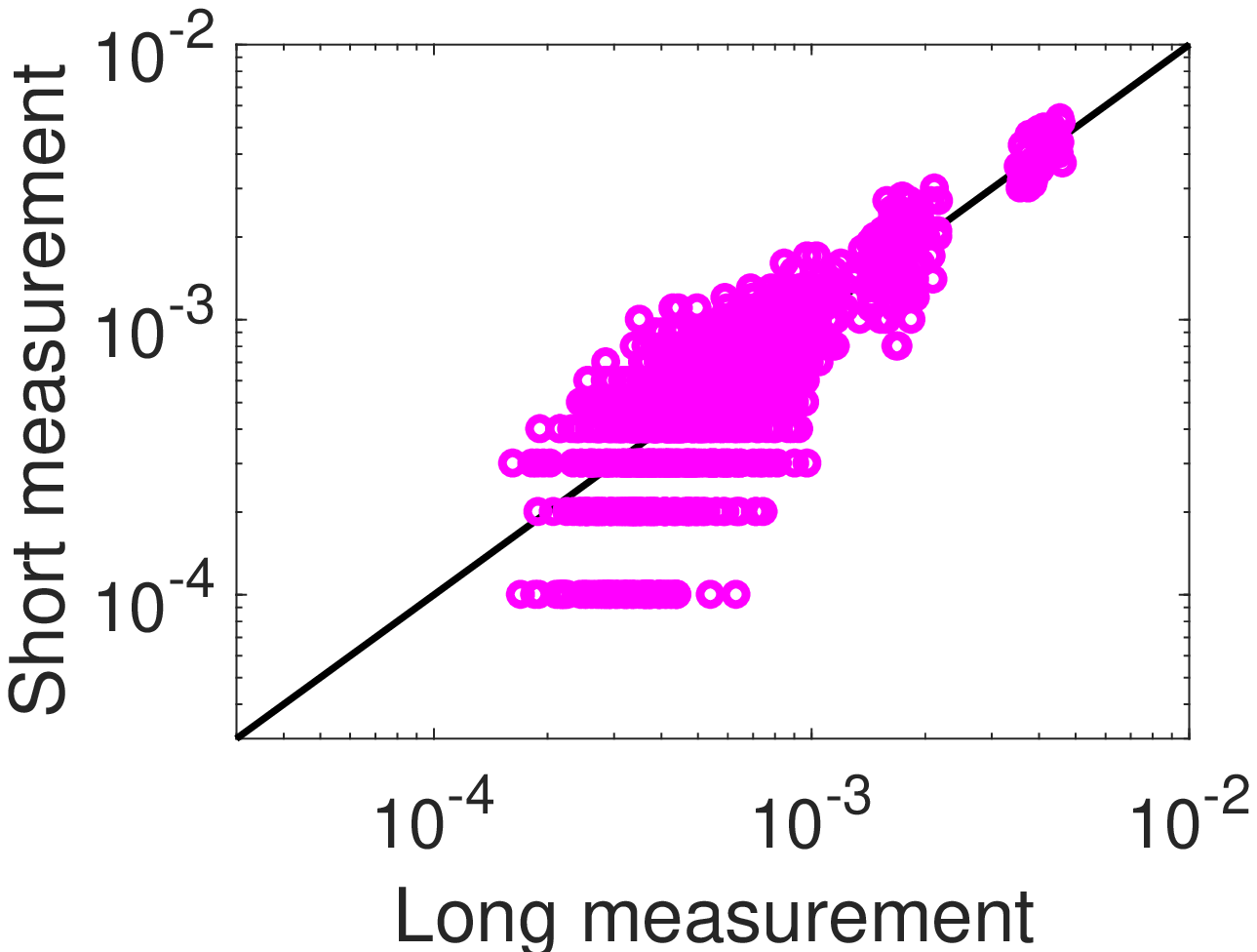} 
\par\end{centering}
}\subfloat[]{\begin{centering}
\includegraphics[scale=0.42]{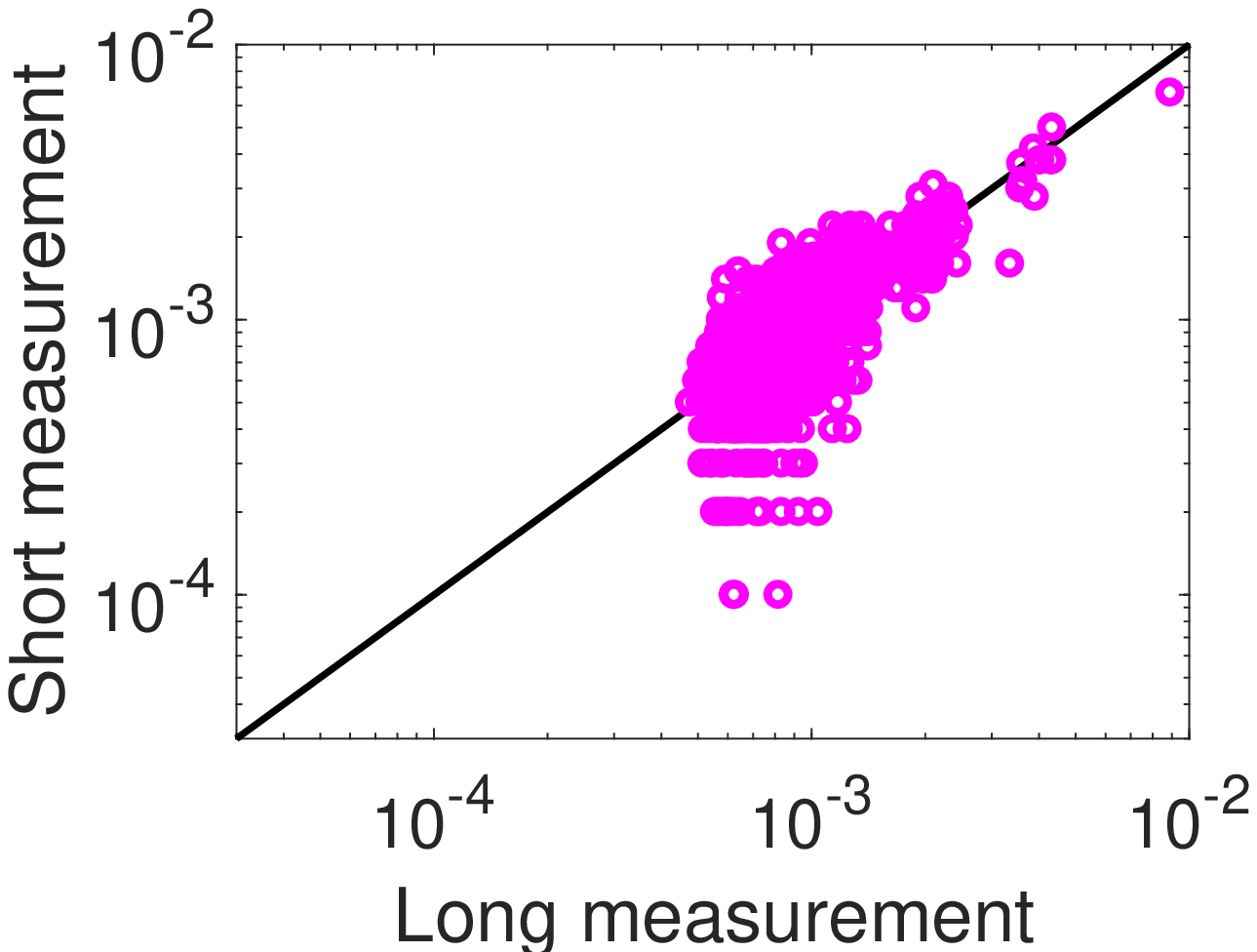} 
\par\end{centering}
}

\caption{\textbf{Measured probability distributions for short-time recordings
compared with long-time recordings.} The frequency of each firing
state measured from the short-time recording of network dynamics ($\unit[120]{s}$)
is plotted against the frequency measured from the long-time recording
of network dynamics ($\unit[1.2\times10^{5}]{s}$). Data for these
three cases are from three dynamical regimes shown (the same $10$
selected neurons in each case) in Fig. \ref{fig:Raster}, respectively.\label{fig:short}}
\end{figure*}
\par\end{center}

\subsection*{One-to-one mapping between the probability distribution and the moments}

To demonstrate the relationship between the true probability distribution
of network states, $P_{{\rm true}}(\Omega)$, and the corresponding
moments, we introduce several notations for ease of discussion. First,
denote the vector $\mathbf{P}^{(n)}=(p_{1}^{(n)},p_{2}^{(n)},\cdots,p_{2^{n}}^{(n)})^{T}$
as the vector containing the probability distribution of the network
states for a network of $n$ nodes, and denote the vector $\mathbf{M}^{(n)}=(m_{1}^{(n)},m_{2}^{(n)},\cdots,m_{2^{n}}^{(n)})^{T}$
as the vector containing all moments of the network. We arrange the
entries in $\mathbf{P}^{(n)}$ and $\mathbf{M}^{(n)}$ as follows.
For an example network of size $n=2$, we assign values to the $i$th
entry by expressing $i-1$ using the base-2 number system with total
$n=2$ digits, i.e., $E_{i}=e_{2}e_{1}$, where $E_{i}$ represents
the combined state of the two nodes in the network, $e_{1}$ and $e_{2}$
(e.g., 00 corresponds to both nodes being inactive), as shown in the
second column of Table \ref{Table:exQ_R}. Then, for each $i$, denote
the probability of the network state ($\sigma_{1}=e_{1},\sigma_{2}=e_{2})$
as $p_{i}^{(2)}$, as shown in the third column of Table \ref{Table:exQ_R}.
In neuroscience, the vector $P_{00}$ represents the probability of
finding both two neurons in the quiet state, $P_{10}$ ($P_{01}$)
represents the probability of the first (second) neuron in the active
state and the second (first) neuron in the silent state, and $P_{11}$
represents the probability of both neurons in the active state. The
entries in the vector $\mathbf{M}^{(2)}$ are arranged similarly,
i.e., $m_{i}^{(2)}$ is the expectation of $(\sigma_{1})^{e_{1}}(\sigma_{2})^{e_{2}}$,
as shown in the fourth column of Table \ref{Table:exQ_R}.

\begin{table}
\begin{centering}
\begin{tabular}{|c|c|c|c|}
\hline 
i  & $E_{i}=e_{2}e_{1}$  & $p_{i}^{(2)}$  & $m_{i}^{(2)}$\tabularnewline
\hline 
\hline 
1  & 00  & $P_{00}$  & 1\tabularnewline
\hline 
2  & 01  & $P_{10}$  & $\left\langle \sigma_{1}\right\rangle $\tabularnewline
\hline 
3  & 10  & $P_{01}$  & $\left\langle \sigma_{2}\right\rangle $\tabularnewline
\hline 
4  & 11  & $P_{11}$  & $\left\langle \sigma_{1}\sigma_{2}\right\rangle $\tabularnewline
\hline 
\end{tabular}
\par\end{centering}
\caption{Table of entries for building the vectors $\mathbf{P}^{(n)}$ and
$\mathbf{M}^{(n)}$ for an example network of $n=2$ nodes.\label{Table:exQ_R}}
\end{table}

For illustration, we show that, for a network of $n=2$ nodes, there
is a full-rank matrix, $\mathbf{U}_{{\rm PM}}^{(2)}$, that transforms
from the probability distribution to moments. From Eqs. (\ref{eq:1moment})
and (\ref{eq:2moment}), the expectation of $\sigma_{1}$, $\sigma_{2}$,
and $\sigma_{1}\sigma_{2}$ can be obtained by summing the probabilities
in which those nodes are active, leading to the following system
\begin{equation}
\left[\begin{array}{cccc}
1 & 1 & 1 & 1\\
0 & 1 & 0 & 1\\
0 & 0 & 1 & 1\\
0 & 0 & 0 & 1
\end{array}\right]\left[\begin{array}{c}
P_{00}\\
P_{10}\\
P_{01}\\
P_{11}
\end{array}\right]=\left[\begin{array}{c}
1\\
\left\langle \sigma_{1}\right\rangle \\
\left\langle \sigma_{2}\right\rangle \\
\left\langle \sigma_{1}\sigma_{2}\right\rangle 
\end{array}\right],\label{eq:N2Cp}
\end{equation}
i.e., $\mathbf{U}_{{\rm PM}}^{(2)}\mathbf{P}^{(2)}=\mathbf{M}^{(2)}$.
Clearly, from Eq. (\ref{eq:N2Cp}), $\mathbf{U}_{{\rm PM}}^{(2)}$
is upper-triangular and of full rank. 

The above analysis can be extended to a network of any size $n$.
For each integer $i$, $1\leq i\leq2^{n}$, we can similarly express
$i-1$ by the base-2 number system with $n$ digits, denoted by $E_{i}=e_{n}e_{n-1}\cdots e_{2}e_{1}$.
Then, write the probability of the network state ($\sigma_{1}=e_{1},\sigma_{2}=e_{2},\cdots,\sigma_{n}=e_{n})$,
denoted as $p_{i}^{(n)}$, and the moment i.e., the expectation of
$(\sigma_{1})^{e_{1}}(\sigma_{2})^{e_{2}}\cdots(\sigma_{n-1})^{e_{n-1}}(\sigma_{n})^{e_{n}}$,
denoted as $m_{i}^{(n)}$, as follows 
\begin{equation}
\mathbf{P}^{(n)}=\left[\begin{array}{c}
P_{00\cdots0}\\
P_{10\cdots0}\\
P_{01\cdots0}\\
P_{11\cdots0}\\
\vdots\\
P_{01\cdots1}\\
P_{11\cdots1}
\end{array}\right],\quad\mathbf{M}^{(n)}=\left[\begin{array}{c}
1\\
\left\langle \sigma_{1}\right\rangle \\
\left\langle \sigma_{2}\right\rangle \\
\left\langle \sigma_{1}\sigma_{2}\right\rangle \\
\vdots\\
\left\langle \prod_{j=2}^{n}\sigma_{j}\right\rangle \\
\left\langle \sigma_{1}\prod_{j=2}^{n}\sigma_{j}\right\rangle 
\end{array}\right].\label{eq:QR}
\end{equation}

We prove that there is a full-rank matrix, ${\bf U}_{{\rm PM}}^{(n)}$,
that transforms from $\mathbf{P}^{(n)}$ to $\mathbf{M}^{(n)}$ as
\begin{equation}
{\bf U}_{{\rm PM}}^{(n)}\mathbf{P}^{(n)}=\mathbf{M}^{(n)},\label{eq:AQR}
\end{equation}
where ${\bf U}_{{\rm PM}}^{(n)}$ is upper-triangular and of full
rank (see Appendix \ref{subsec:A3:-The-proof} for details).

Therefore, all moments, which are shown in the entries of $\mathbf{M}^{(n)}$,
can be used to describe the probability distribution of network states
for a network of $n$ nodes with binary dynamics. As the full-order
MEP distribution, $P_{n}(\Omega)$, is subject to all moments, $P_{n}(\Omega)$
and $P_{{\rm true}}(\Omega)$ share the same moments for a sufficiently
long-time recording, i.e., $\mathbf{M}^{(n)}$ in Eq. (\ref{eq:AQR}).
Since ${\bf U}_{{\rm PM}}^{(n)}$ is of full rank, $P_{n}(\Omega)$
is identical to $P_{{\rm true}}(\Omega)$.

By directly substituting $P_{{\rm true}}(\Omega)$ into the full-order
MEP analysis, we develop a relationship between effective interactions
and the probability distribution, as discussed in the next section.

\subsection*{One-to-one mapping between effective interactions and the probability
distribution }

To demonstrate the relationship between effective interactions and
the true probability distribution of network states, we substitute
all $2^{n}$ states of $\Omega=(\sigma_{1},\sigma_{2},\cdots,\sigma_{n})$
and the probability distribution, $P_{{\rm true}}(\Omega)$, into
Eq. (\ref{eq:PV}) with $m=n$, and then take the logarithm of both
sides. This results in a system of linear equations in terms of $-\log Z$
and all the effective interactions, 
\begin{equation}
-\log Z+\sum_{k=1}^{n}\sum_{i_{1}<\cdots<i_{k}}^{n}J_{i_{1}\cdots i_{k}}\sigma_{i_{1}}\cdots\sigma_{i_{k}}=\log P_{{\rm true}}(\Omega),\label{eq:SolveJ3}
\end{equation}
where $-\log Z$ can be regarded as the zeroth-order effective interaction,
$J_{0}$. By solving the system of linear equations in Eq. (\ref{eq:SolveJ3}),
we can obtain all the $2^{n}$ effective interactions, $J$'s, in
terms of the true probability distribution of network states, $P_{{\rm true}}(\Omega)$.

We again turn to the small network case of $n=2$ nodes to demonstrate
how to obtain a one-to-one mapping from the linear system described
by Eq. (\ref{eq:SolveJ3}). First, denote the vector $\mathbf{J}^{(2)}$
as the vector containing all the effective interactions, with the
index of each effective interaction, $i$. We then express $i-1$
by the base-2 number system with total $n=2$ digits, denoted by $E_{i}=e_{2}e_{1}$,
and the $i$th entry of $\mathbf{J}^{(2)}$ is the coefficient of
the term $(\sigma_{1})^{e_{1}}(\sigma_{2})^{e_{2}}$ in Eq. (\ref{eq:SolveJ3}),
yielding $\mathbf{J}^{(2)}=(J_{0},\,J_{1},\,J_{2},\,J_{12})^{T}$.
Since the ordering of the indices for $\mathbf{J}^{(2)}$ is the same
as that of $\mathbf{P}^{(2)}$, then the right hand side of Eq. (\ref{eq:SolveJ3})
is simply the logarithm of the vector $\mathbf{P}^{(2)}$. Therefore,
for a network of $n=2$ nodes, we have the following equation

\begin{equation}
\left[\begin{array}{cccc}
1 & 0 & 0 & 0\\
1 & 1 & 0 & 0\\
1 & 0 & 1 & 0\\
1 & 1 & 1 & 1
\end{array}\right]\left[\begin{array}{c}
J_{0}\\
J_{1}\\
J_{2}\\
J_{12}
\end{array}\right]=\left[\begin{array}{c}
\log P_{00}\\
\log P_{10}\\
\log P_{01}\\
\log P_{11}
\end{array}\right].\label{BJQ2}
\end{equation}

The above relation can be extended to a network of any size $n$ and
the corresponding linear equations are as follows 
\begin{equation}
{\bf L}_{{\rm JP}}^{(n)}\mathbf{J}^{(n)}=\log\mathbf{P}^{(n)},\label{eq:BJQ}
\end{equation}
where ${\bf L}_{{\rm JP}}^{(n)}$ is a lower-triangular matrix with
dimension $2^{n}\times2^{n}$. For $1\leq i\leq2^{n}$, we can similarly
express $i-1$ by the base-2 number system with $n$ digits, denoted
by $E_{i}=e_{n}e_{n-1}\cdots e_{2}e_{1}$. Then, the $i$th entry
of $\mathbf{J}^{(n)}$ is the coefficient of the term $(\sigma_{1})^{e_{1}}(\sigma_{2})^{e_{2}}\cdots(\sigma_{n-1})^{e_{n-1}}(\sigma_{n})^{e_{n}}$
in Eq. (\ref{eq:SolveJ3}). To show the linear transform between the
effective interactions in the full-order MEP distribution $\mathbf{J}^{(n)}$
and the probability distribution of network states $\mathbf{P}^{(n)}$
is a one-to-one mapping, one needs to demonstrate that ${\bf L}_{{\rm JP}}^{(n)}$
is a matrix of full rank. Similarly, as the proof of the one-to-one
mapping between the probability distribution and the moments, we can
show that ${\bf L}_{{\rm JP}}^{(n)}$ is the transpose of ${\bf U}_{{\rm PM}}^{(n)}$,
i.e., ${\bf L}_{{\rm JP}}^{(n)}=\left(\mathbf{U}_{{\rm PM}}^{(n)}\right)^{T}$
by mathematical induction. Therefore, ${\bf L}_{{\rm JP}}^{(n)}$
is lower-triangular and full-rank, and one can use effective interactions
to characterize the probability distribution of network states for
a network of nodes with binary dynamics.

\subsection*{High-order effective interactions are small for asynchronous networks}

Using the mapping between the probability distribution and effective
interactions, we show that there is a recursive relationship among
different orders of the effective interactions, where high-order effective
interactions are weak compared with low-order ones in asynchronous
networks. For illustration, we first discuss the case of a network
with size $n=2$. Based on Eq. (\ref{BJQ2}), we have
\[
J_{0}=\log P_{00},\quad J_{1}=\log\frac{P_{10}}{P_{00}},\quad J_{2}=\log\frac{P_{01}}{P_{00}},
\]
and 
\[
J_{12}=\log\frac{P_{11}}{P_{01}}-\log\frac{P_{10}}{P_{00}}.
\]
Next, we define a new term 
\[
J_{1}^{1}\triangleq\log\frac{P_{11}}{P_{01}},
\]
which describes the case in which the state of the second node in
the network is changed from inactive (state $0$) to active (state
$1$) (i.e., in $J_{1}$, $P_{10}\rightarrow P_{11}$ and $P_{00}\rightarrow P_{01}$).
Note that with this notation, we can now express the higher-order
effective interaction, $J_{12}$, as 
\[
J_{12}=J_{1}^{1}-J_{1}.
\]
 For a network of any size $n$, based on Eq. (\ref{eq:BJQ}), we
obtain an expression for the first-order effective interaction 
\begin{equation}
J_{1}=\log\frac{P_{10\cdots0}}{P_{00\cdots0}}\label{eq:hP}
\end{equation}
and for the second-order effective interaction 
\begin{equation}
J_{12}=\log\frac{P_{110\cdots0}}{P_{010\cdots0}}-\log\frac{P_{10\cdots0}}{P_{00\cdots0}}.\label{eq:J12Theo}
\end{equation}
Then, the second-order effective interaction, $J_{12}$, can be equivalently
obtained by the following procedure: First, in $J_{1}=\log(P_{10\cdots0}/P_{00\cdots0}),$
we switch the state of the second node from 0 to 1 to obtain a new
term $J_{1}^{1}=\log(P_{110\cdots0}/P_{010\cdots0})$. Then, note
that if we subtract $J_{1}$ from $J_{1}^{1}$, we arrive at $J_{12}$
as described in Eq. (\ref{eq:J12Theo}). The above notation can be
extended to the case of high-order effective interactions (see Appendix
\ref{sec:Proof-of-Recursive} for a proof): 
\begin{equation}
J_{12\cdots(k+1)}=J_{12\cdots k}^{1}-J_{12\cdots k},\label{eq:Jobtain}
\end{equation}
where $1\leq k\leq n-1$ and $J_{12\cdots k}^{1}$ is obtained by
switching the state of the $(k+1)$st node in $J_{12\cdots k}$ from
0 to 1. Next, we intuitively explain that the recursive structure
(Eq. (\ref{eq:Jobtain})) leads to the hierarchy of effective interactions,
i.e., high-order effective interactions are much smaller than low-order
ones for asynchronous networks. $J_{12\cdots k}$ describes the effective
interaction of the sub-network of neurons $\{1,2,3,...,k\}$ when
the state of the $(k+1)$st neuron and states of neurons with indices
$\{k+2,k+3,...,n\}$ are inactive. The term, $J_{12\cdots k}^{1}$,
describes the case in which the $(k+1)$st neuron becomes active and
states of all other neurons are not changed. For an asynchronous network,
the cortical interactions between neurons are relatively weak, therefore,
the effect on the change of network dynamics from the case when the
$(k+1)$st neuron is silent to the case when the $(k+1)$st neuron
is active should be small. In other words, the value difference between
the effective interaction of the sub-network where the $(k+1)$st
neuron is inactive, $J_{12\cdots k}$, and the effective interaction
when the $(k+1)$st neuron is active, $J_{12\cdots k}^{1}$, should
be small, and thus the higher-order effective interaction, $J_{12\cdots(k+1)}$,
is much smaller than the low-order effective interaction, $J_{12\cdots k}$.
Note that more strict mathematical proof of the hierarchy of effective
interactions in an asynchronous network can be seen in our previous
work \cite{xu2016dynamical}. Therefore, the recursive relation in
Eq. (\ref{eq:Jobtain}) leads to a hierarchy of effective interactions,
i.e., low-order effective interactions dominate higher-order effective
interactions in the MEP analysis. 

Next, we investigate the HH neuronal network dynamics to confirm that
the first two-order effective interactions in the full-order MEP distribution
dominate higher-order effective interactions. We evolve HH networks
in a long-time simulation of $\unit[1.2\times10^{5}]{s}$ and measure
the probability distribution of network states. We have verified that
the HH network dynamics in such a long simulation time reaches the
steady state, therefore, the measured probability distribution of
network states can well represent the true probability distribution
of network states. We then calculate effective interactions in the
full-order MEP distribution by Eq. (\ref{eq:BJQ}) from the measured
probability distribution of network states. In Fig. \ref{fig:Jplot},
the average strength of the $k$th-order effective interactions is
computed as the mean of the absolute value of the $k$th-order effective
interactions. It can clearly be seen that for three different dynamical
regimes, the average strength of effective interactions of high-orders
($\geq3$) are at least one order of magnitude smaller (in the absolute
value) than that of the first-order and the second-order effective
interactions.
\begin{center}
\begin{figure*}
\subfloat[]{\begin{centering}
\includegraphics[scale=0.42]{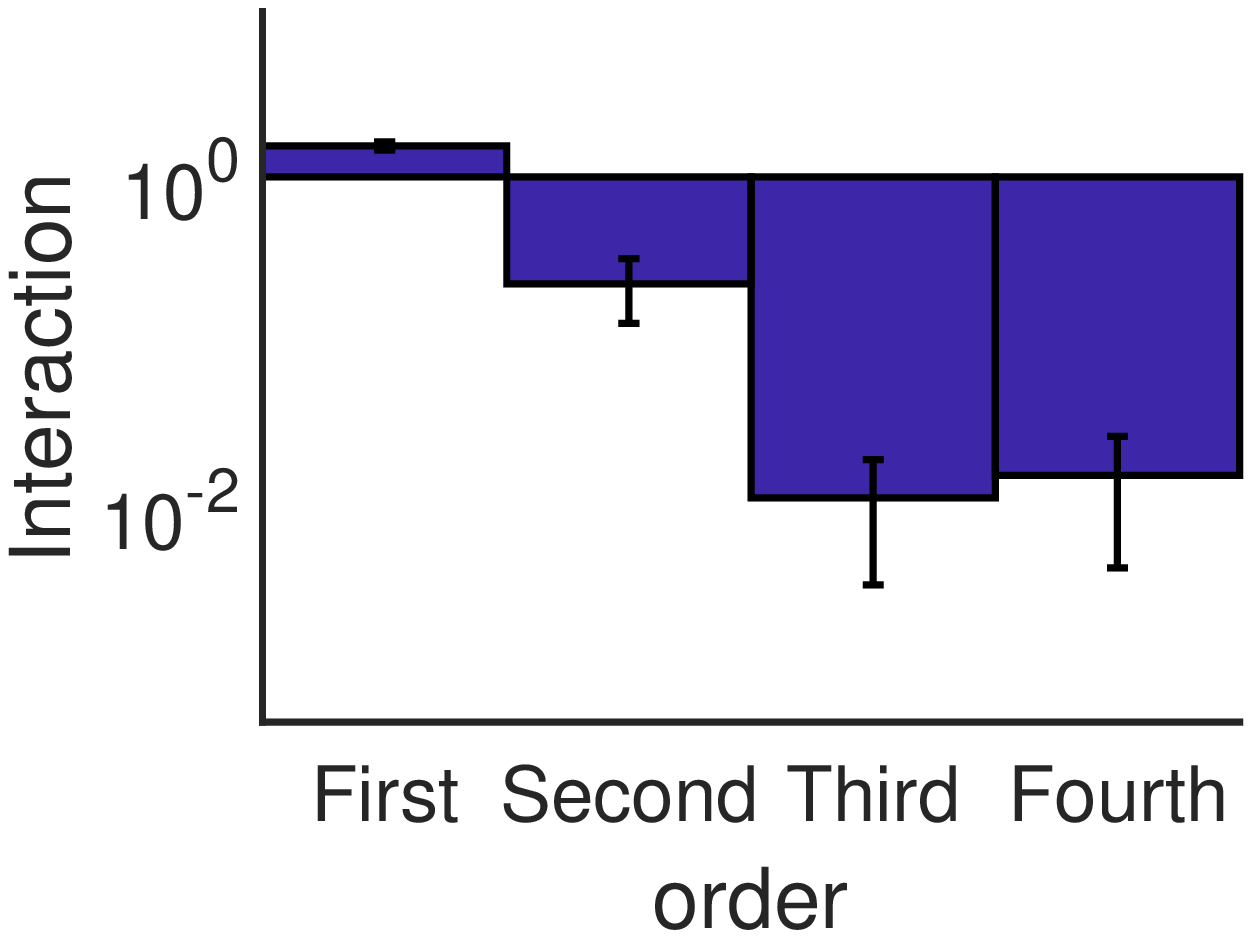} 
\par\end{centering}
}\subfloat[]{\begin{centering}
\includegraphics[scale=0.42]{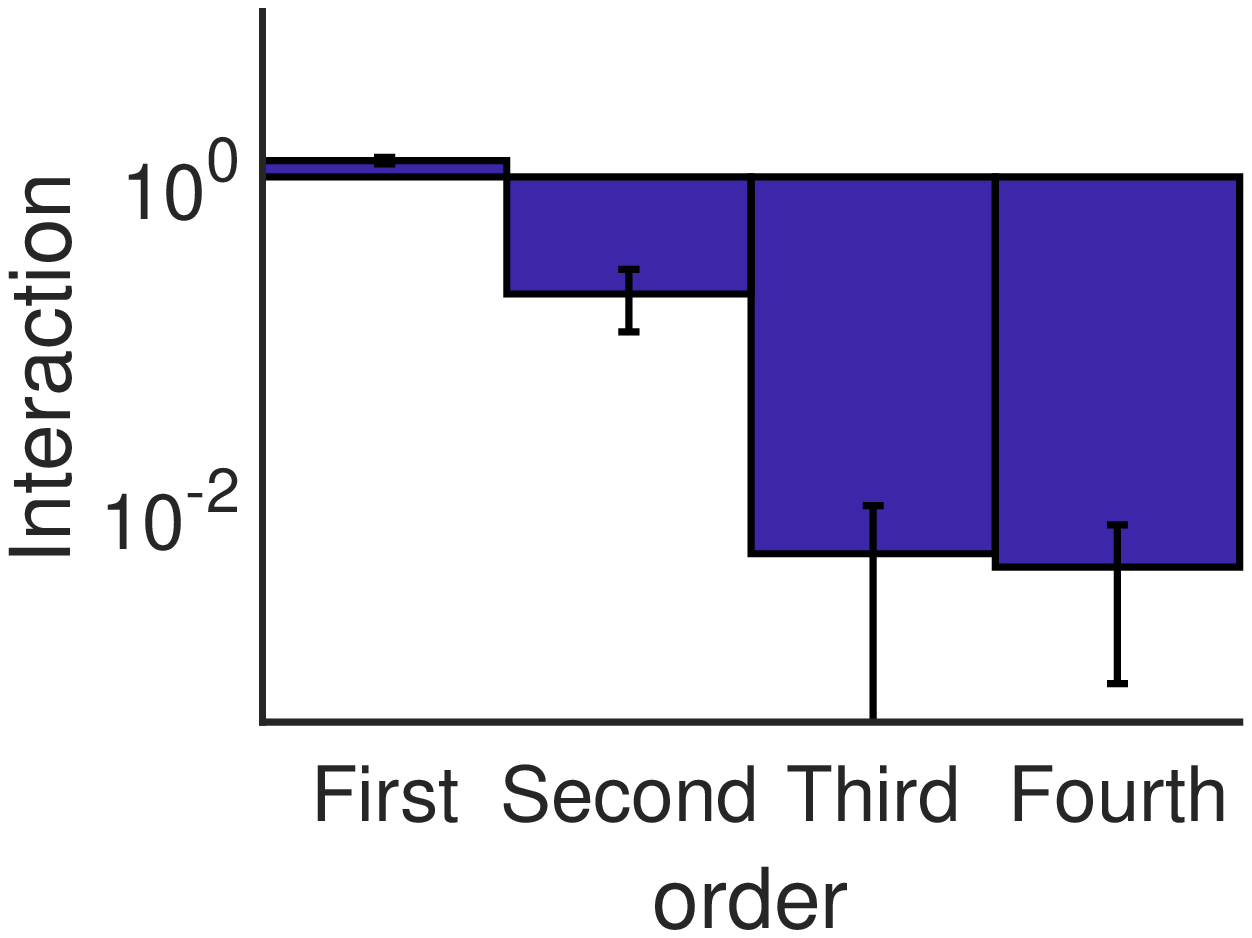} 
\par\end{centering}
}\subfloat[]{\begin{centering}
\includegraphics[scale=0.42]{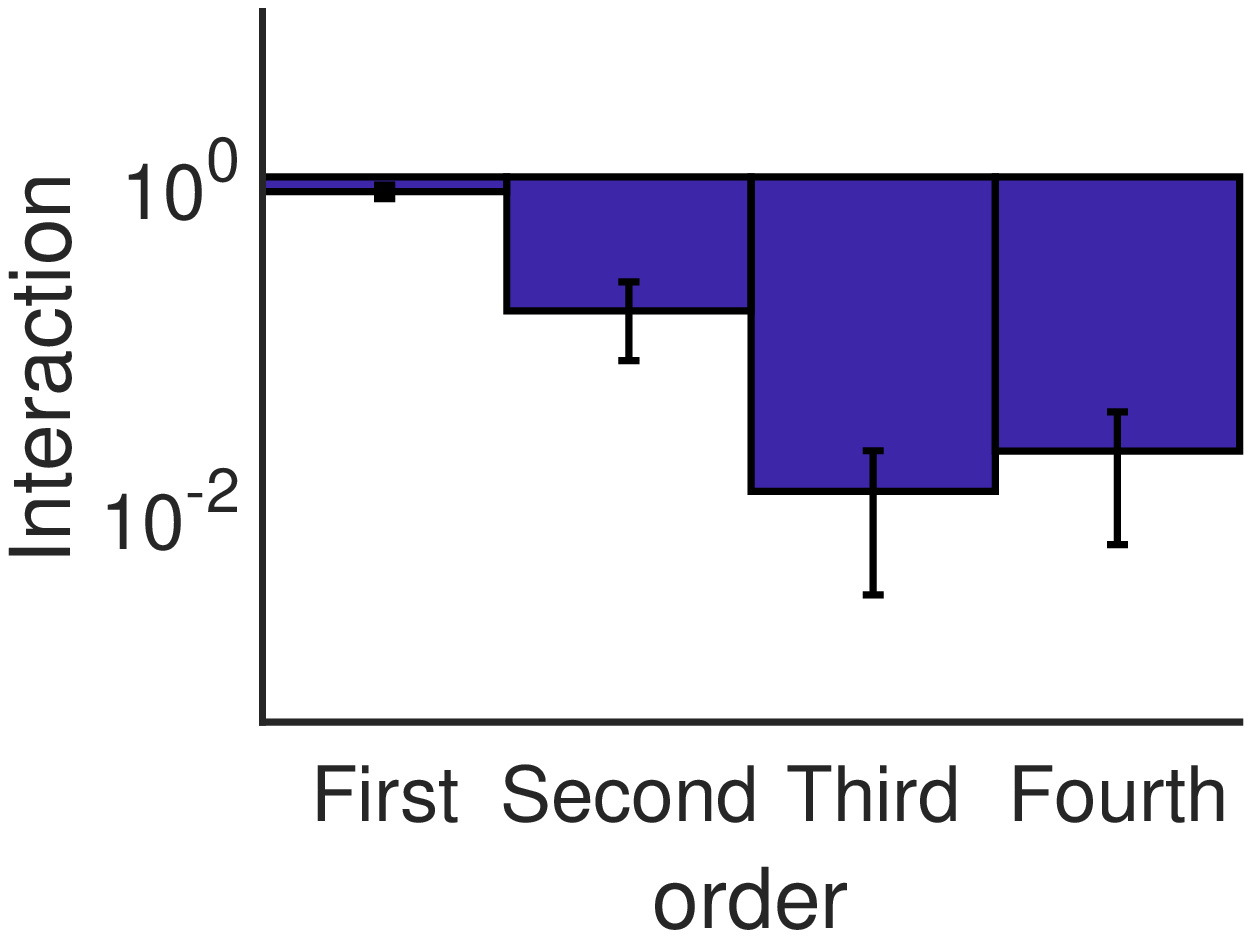} 
\par\end{centering}
}

\caption{\textbf{Hierarchy of effective interactions in the MEP analysis in
three dynamical regimes.} Mean absolute values of effective interactions
from the first-order to the fourth-order for three different dynamical
regimes (shown in Fig. \ref{fig:Raster}) are plotted, with the standard
deviation indicated by the error bar. Data for these three cases are
collected from the three experiments (the same $10$ selected neurons
in each case) in Fig. \ref{fig:Raster}, respectively, with a long
recording time of $\unit[1.2\times10^{5}]{s}$. \label{fig:Jplot}}
\end{figure*}
\par\end{center}

\subsection*{Low-order MEP analysis with short-time recordings}

Using the mappings described in the previous sections, we show that
the low-order MEP analysis can provide an accurate estimate of the
probability distribution of network states with a short-time recording.

First, note that in a short recording of an asynchronous network,
the probability of observing a highly-active state (one in which many
nodes are active simultaneously) is usually too small to be measured
accurately. Thus, the probability distribution of network states measured
from the short-time recording cannot capture many network activity
states as observed in the long-time recordings (e.g., shown in Fig.
\ref{fig:short}). Note that the mapping between the probability distribution
and effective interactions, i.e., the lower-triangular matrix ${\bf L}_{{\rm JP}}^{(n)}$,
suggests that the probability of a highly-active state consists of
the summation of many effective interactions. For example, in a small
network of $n=2$ nodes, the probability of the highly-active state,
$P(\sigma_{1}=1,\sigma_{2}=1)$, can be obtained from Eq. (\ref{BJQ2})
by: $\log P_{11}=J_{0}+J_{1}+J_{2}+J_{12}$. Therefore, to obtain
an accurate estimation of such highly-active states, it is important
to obtain an accurate estimation of the summation of the effective
interactions.

For asynchronous networks, since high-order effective interactions
are small (e.g., shown in Fig. \ref{fig:Jplot}), the summation in
the probability of a highly-active state will be dominated by low-order
effective interactions. Note that low-order effective interactions
can be derived from low-order moments through an iterative scaling
algorithm (see Appendix \ref{subsec:A2:-The-iterative} for details)
in the MEP analysis, and thus an accurate estimation of low-order
moments is essential. The mapping between the probability distribution
of network states and corresponding moments of the network, i.e.,
the upper triangular matrix ${\bf U}_{{\rm PM}}^{(n)}$ (Eq. (\ref{eq:N2Cp})),
shows that low-order moments consist of the summation of probabilities
of many network activity states. For example, in a network of $n$
nodes, the first-order moment of node 1 is the summation of network
state probabilities where node 1 is active (number of $2^{n-1}$ states
in total from low-active states to highly-active states), that is
$\left\langle \sigma_{1}\right\rangle =P_{10\cdots0}+P_{11\cdots0}+\cdots+P_{11\cdots1}$.
Note that low-active states occurs frequently in an asynchronous network;
therefore, the summation in low-order moments is dominated by probabilities
of these low-active states which can be accurately measured from a
short-time recording, leading to good estimation of low-order moments.
In addition, we point out that the estimation error in low-order moments
can be further reduced due to linear summations. Therefore, one can
accurately estimate the low-order moments and perform the low-order
MEP analysis in a short-time data recording of network dynamics. It
is expected that the low-order MEP distribution provides a good estimate
of the probability distribution of network states.

In summary, the low-order MEP distribution can be obtained by the
following procedure: First, calculate low-order moments from the experimental
short-time recording using Eqs. (\ref{eq:1moment}) and (\ref{eq:2moment})
for the first-order and the second-order moments, respectively. Second,
the low-order MEP analysis (with constraints of low-order moments)
is carried out using the iterative scaling algorithm (see Appendix
\ref{subsec:A2:-The-iterative} for details) to derive the low-order
effective interactions with all higher-order effective interactions
set to zero. This determines $\mathbf{J}^{(n)}$ in Eq. (\ref{eq:BJQ}),
except for $J_{0}$. Third, express the probability distribution,
$\mathbf{P}^{(n)}$, as a function of $J_{0}$ using Eq. (\ref{eq:BJQ}).
Finally, determine $J_{0}$ by constraining the summation of all probabilities
to equal one. Once all the low-order effective interactions are determined,
Eq (\ref{eq:PV}) is used to determine the low-order MEP distribution,
e.g., $P_{1}(\Omega)$ and $P_{2}(\Omega)$ represents for the first-order
and the second-order MEP distribution, respectively.

\subsection*{Verification for the MEP analysis with short-time recordings by HH
neuronal network model }

In this section, we use data from the HH neuronal network model to
verify that low-order MEP analysis can accurately estimate the probability
distribution of network states from short-time recordings using the
procedure described in the previous section.

We first evolve the HH neuronal network model with a short simulation
time of $\unit[120]{s}$ and record the spike trains of the network.
Then we estimate the first-order and the second-order moments from
this short recording and use these two-order moments to estimate both
the first-order and the second-order MEP distributions (Eq. (\ref{eq:PV})).
Finally, we compare the MEP distributions to the probability distribution
of network states measured from the long recording of $\unit[1.2\times10^{5}]{s}$.
As shown in Fig. \ref{fig:MEPdis}, for all three dynamical regimes,
the probability distribution of the first-order MEP analysis, $P_{1}$
(green), deviates substantially from the measured probability distribution
of network states in the long recording. The probability distribution
of the second-order MEP analysis, $P_{2}$ (blue), however, is in
excellent agreement with the measured probability distribution of
network states in the long recording. These results indicate that
the low-order, i.e., second-order, MEP analysis provides an efficient
method to obtain the probability distribution of network states with
short recordings.
\begin{center}
\begin{figure*}
\subfloat[]{\begin{centering}
\includegraphics[scale=0.42]{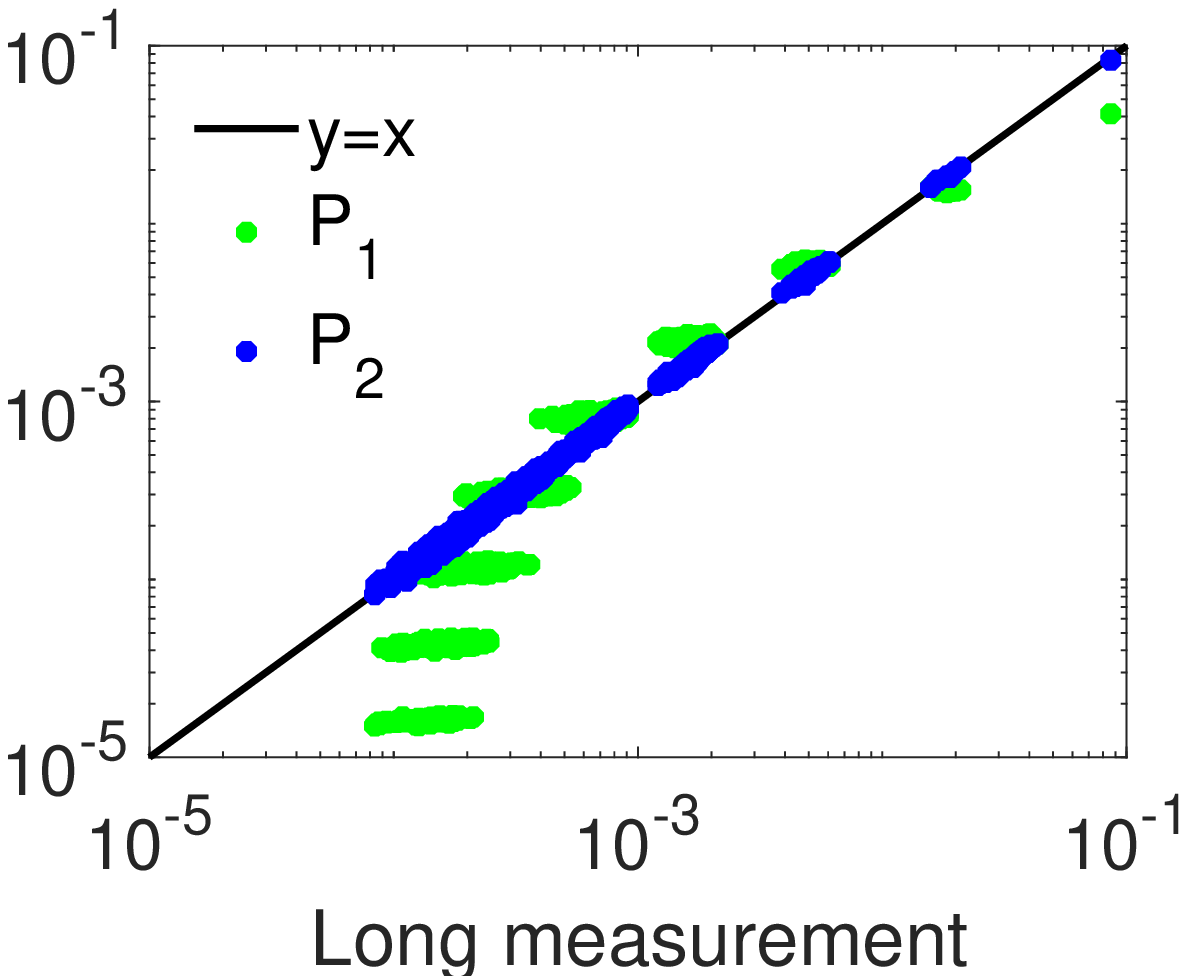} 
\par\end{centering}
}\subfloat[]{\begin{centering}
\includegraphics[scale=0.42]{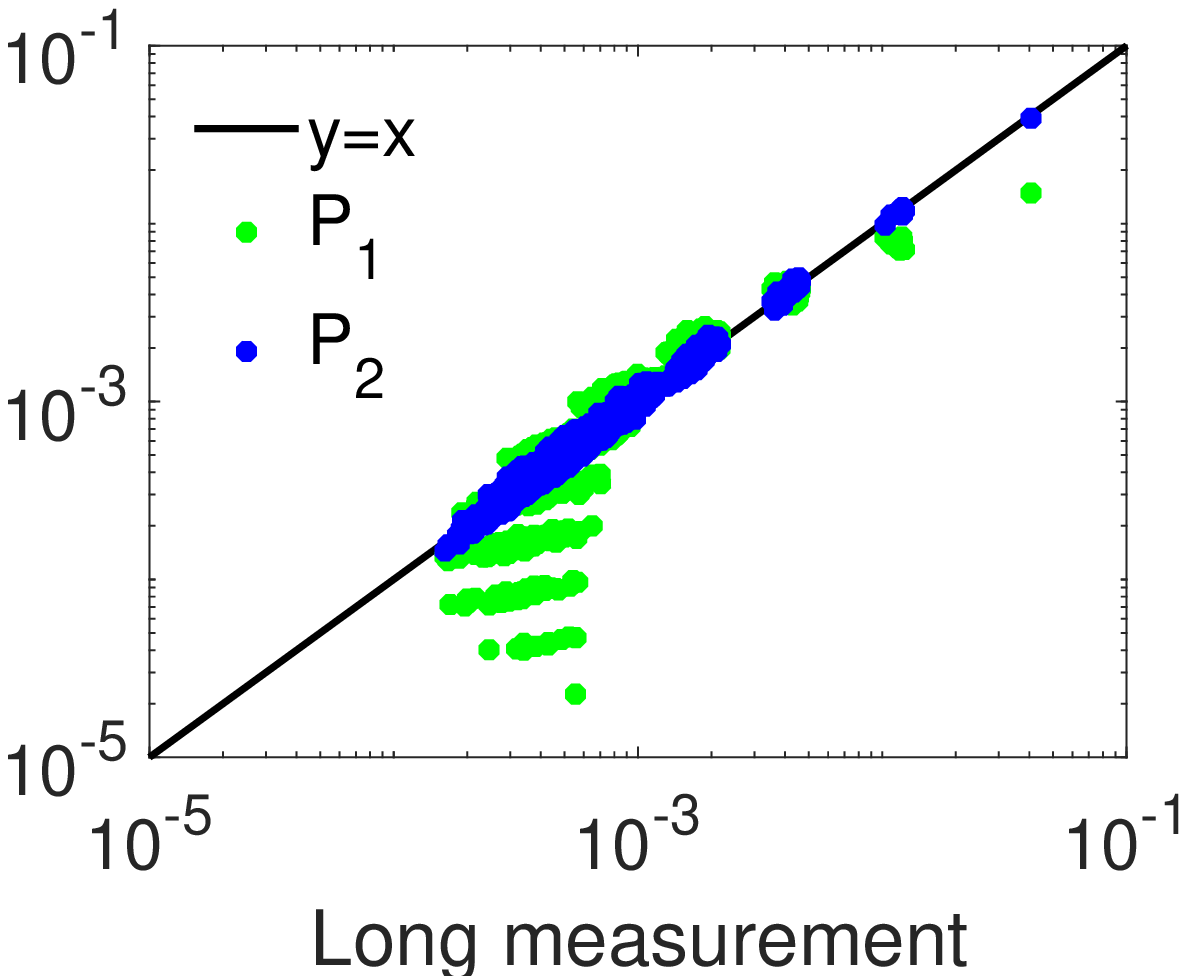} 
\par\end{centering}
}\subfloat[]{\begin{centering}
\includegraphics[scale=0.42]{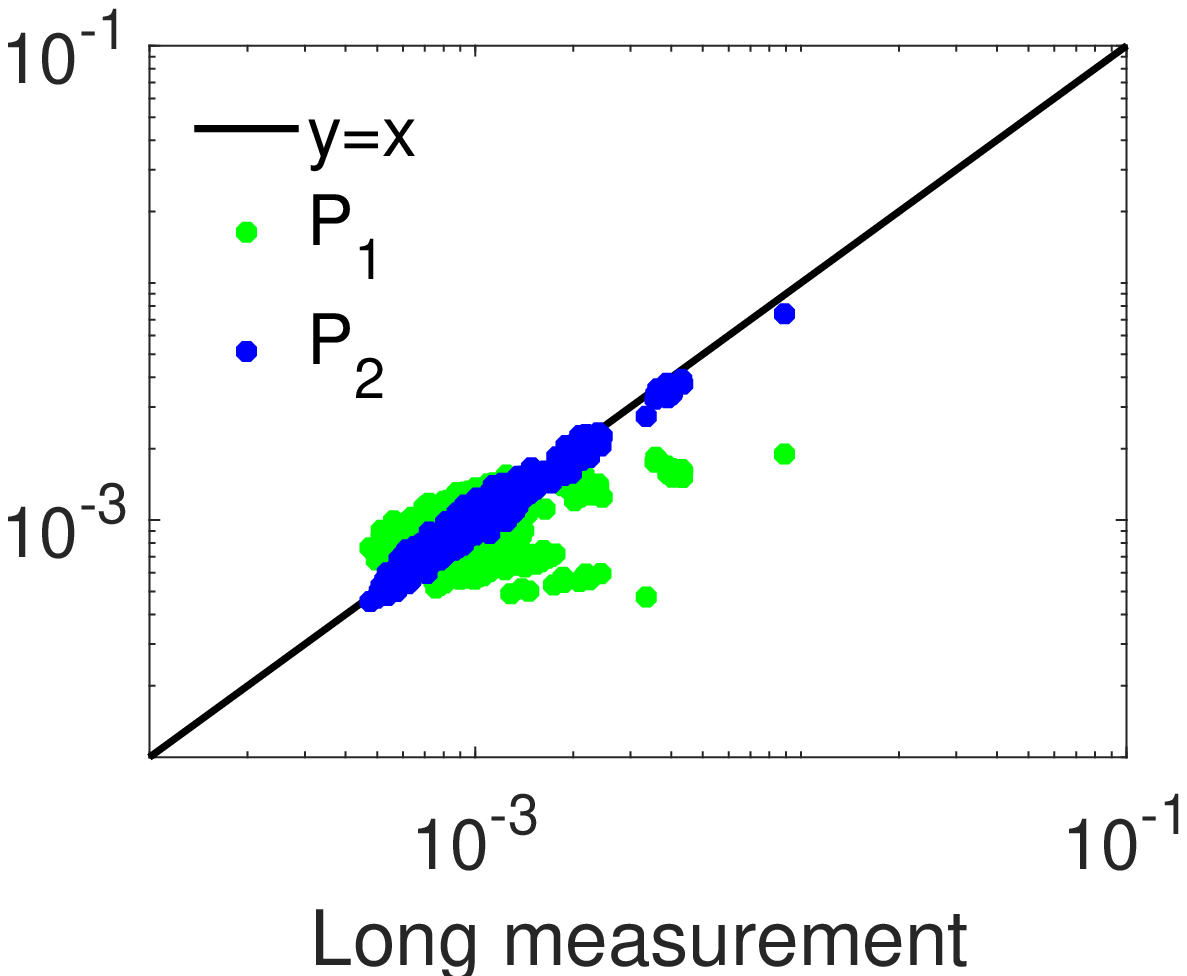} 
\par\end{centering}
}

\caption{\textbf{First-order and second-order MEP distribution in comparison
with the probability distribution of network states.} The frequency
of each firing state from the distribution of the MEP analysis, $P_{2}$
(blue) and $P_{1}$ (green), is plotted against the frequency measured
from the long recording ($\unit[1.2\times10^{5}]{s}$). Data for three
different dynamical regimes (shown in Fig. \ref{fig:Raster}) are
collected from the three experiments (the same $10$ selected neurons
in each case) in Fig. \ref{fig:Raster}, respectively. Here, the MEP
analysis is performed with a short recording ($\unit[1.2\times10^{2}]{s}$).
\label{fig:MEPdis}}
\end{figure*}
\par\end{center}

\subsection*{Verification for the MEP analysis with short-time recordings by electrophysiological
experiments\label{subsec:ExpVerifyMEP}}

In general, it is difficult to obtain a stationary long-time recording
of spike trains from the brain of an awake animal \emph{in vivo}.
To verify the validity of the MEP analysis on short recordings, we
use the electrophysiological experimental data recorded by multi-electrode
array from V1 in \emph{anesthetized} macaque monkeys in multiple trials.
For each trial, an image stimulus was shown on the screen for $\unit[100]{ms}$,
followed by a $\unit[200]{ms}$ uniform gray screen \cite{coen2015flexible,Kohn2015Multi}.
The number of different images is 956 in total and images are presented
in pseudo-random order with each presented $20$ times. Experimental
details can be found in Appendix \ref{sec:pvc8Exp}. Here, we focus
on the issue of whether the second-order MEP analysis from a short
recording can accurately estimate the probability distribution of
neuronal firing patterns in a long recording. Note that the presenting
duration for each image is only $\unit[2]{s}$, which is too short
for the recorded spike trains to have a stable probability distribution.
As an alternative, we put the spike trains recorded during the uniform
gray screen ($\unit[200]{ms}$ in each trial) altogether to obtain
a long recording of $\unit[3824]{s}$. We have verified that the spike
trains in such a long recording has a stable probability distribution.
For a short recording, we randomly select $5\%$ length of the long
recording, i.e., $\unit[191.2]{s}$, and also verified that the probability
distribution in such a short recording is quite different from that
in the long recording. To perform the MEP analysis, we randomly selected
eight neurons' spike train data from experimental measurements. 

As shown in Fig. \ref{fig:MEPExp}a, these eight neurons are in an
asynchronous state. We then calculate effective interactions in the
full-order MEP distribution by Eq. (\ref{eq:BJQ}) using the measured
probability distribution of network states from the long recording
of $\unit[3824]{s}$. In Fig. \ref{fig:MEPExp}b, the average strength
of the $k$th-order effective interactions is computed as the mean
of the absolute value of the $k$th-order effective interactions.
It can be seen clearly that the average strength of effective interactions
of high-orders ($\geq3$) are almost one order of magnitude smaller
(in the absolute value) than that of the first-order and the second-order
effective interactions. Next, we consider a short recording of $\unit[191.2]{s}$
and estimate the first-order and the second-order moments from this
short recording and use these moments to estimate the second-order
MEP distributions (Eq. (\ref{eq:PV})). As shown in Fig.\ref{fig:MEPExp}c,
the probability distribution estimated by the second-order MEP analysis
is in excellent agreement with the measured probability distribution
of network states in the long recording. However, the probability
distribution measured in the short recording clearly deviates from
the probability distribution measured in the long recording. 
\begin{center}
\begin{figure*}
\subfloat[]{\begin{centering}
\includegraphics[scale=0.42]{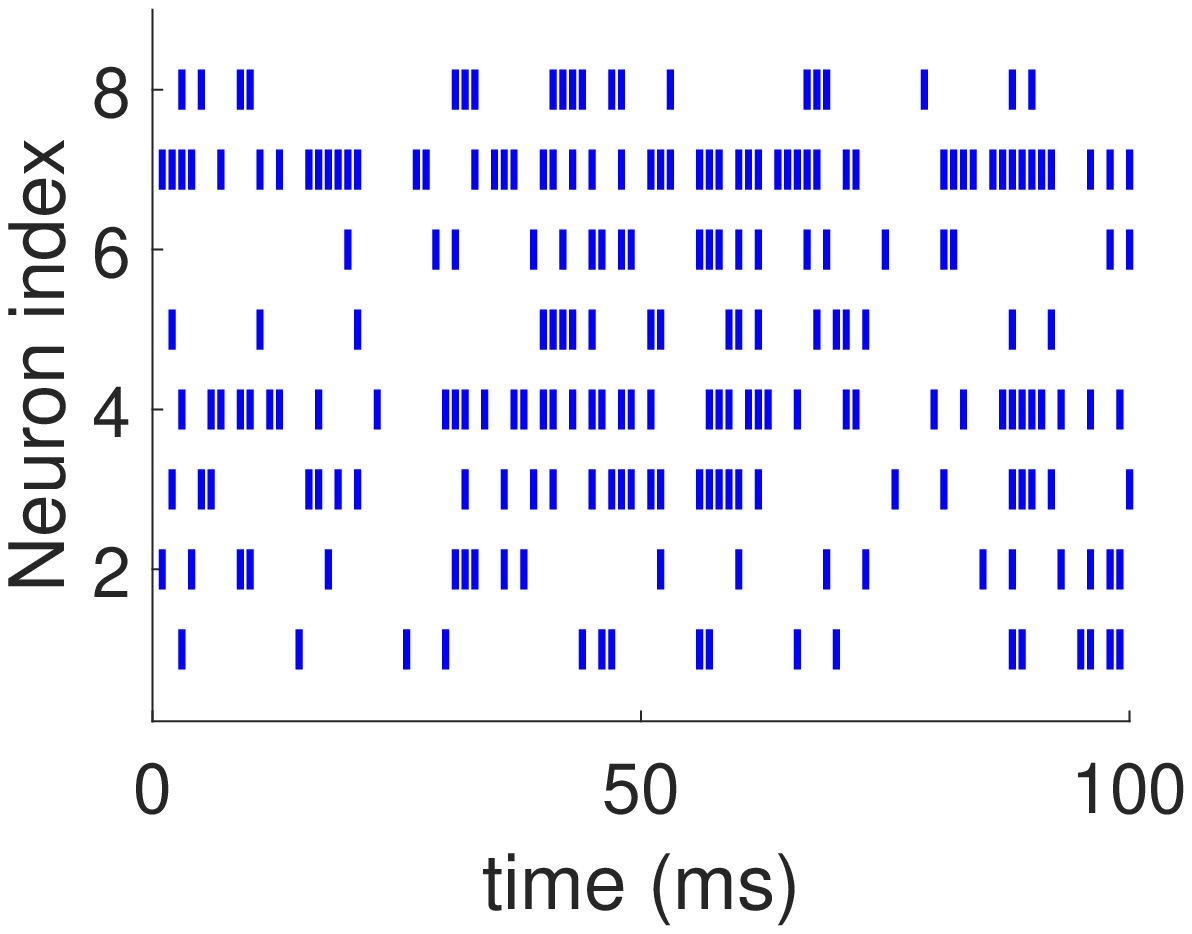} 
\par\end{centering}
}\subfloat[]{\begin{centering}
\includegraphics[scale=0.42]{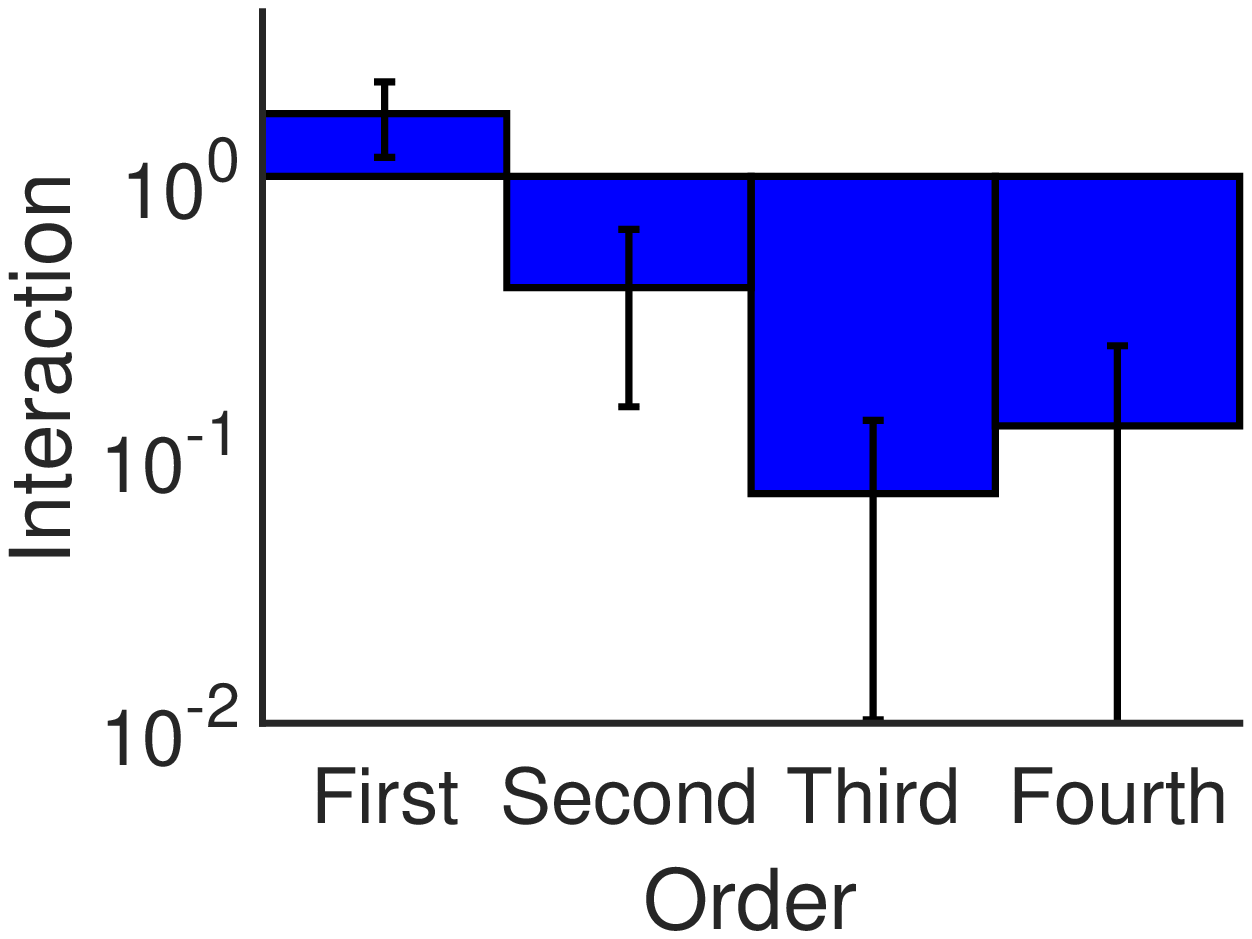} 
\par\end{centering}
}\subfloat[]{\begin{centering}
\includegraphics[scale=0.42]{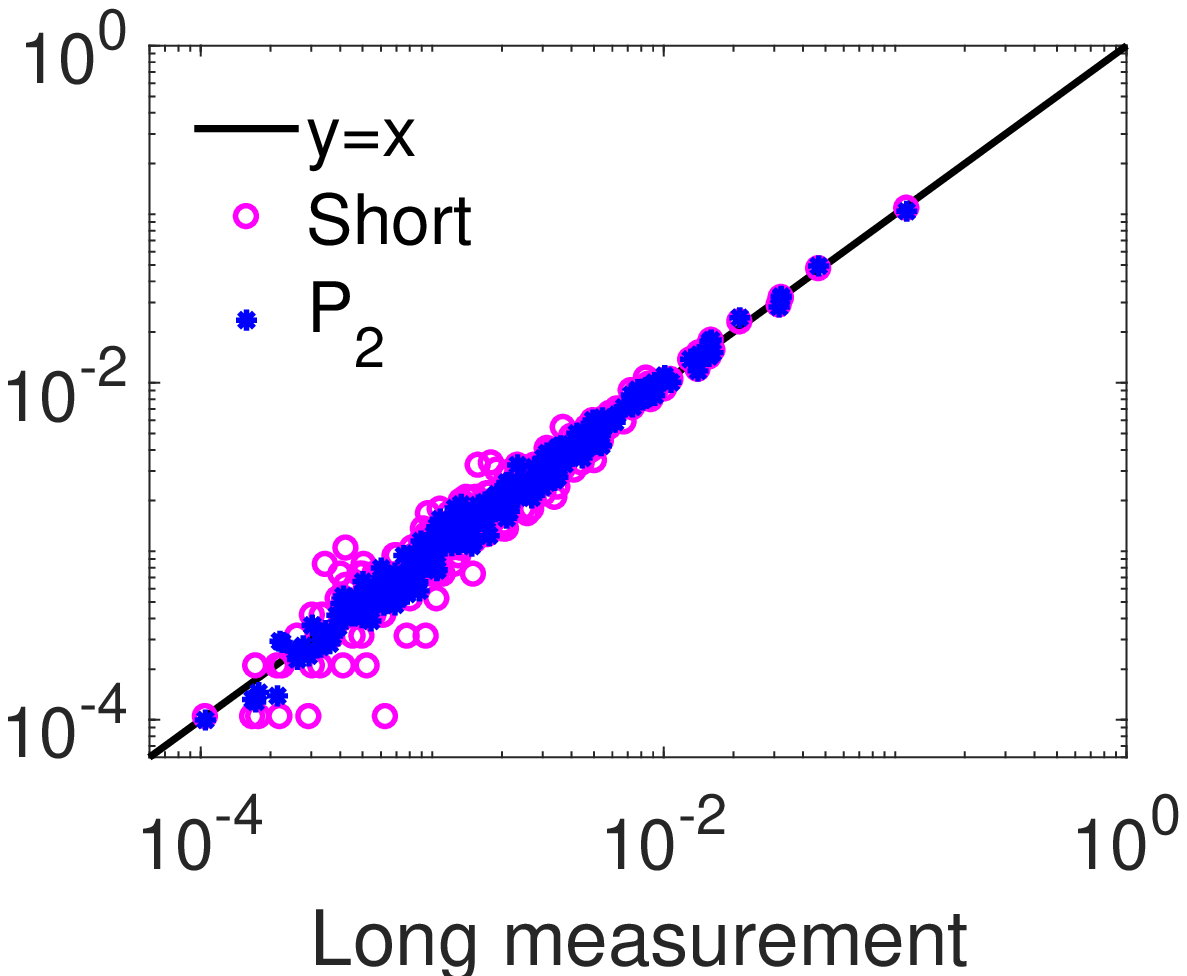} 
\par\end{centering}
}

\caption{\textbf{Electrophysiological verification for the second-order MEP
analysis with short-time recordings.} Spike trains of randomly selected
eight neurons are recorded from V1 in \emph{anesthetized} macaque
monkeys (see Appendix \ref{sec:pvc8Exp}). (a) A short bar indicates
that the neuron with\textcolor{red}{{} }certain index fires at certain
time. (b) Mean absolute values of effective interactions from the
first-order to the fourth-order are plotted, with the standard deviation
indicated by the error bar, computed by the long recording of $\unit[3824]{s}$.
(c) The frequency of each firing state from the distribution measured
in the short recording of $\unit[191.2]{s}$ (magenta) and the distribution
of the second order MEP analysis, $P_{2}$ (blue), is plotted against
the frequency measured from the long recording of $\unit[3824]{s}$.
\label{fig:MEPExp}}
\end{figure*}
\par\end{center}

\section*{Discussion}

The second-order MEP analysis has been used to infer the probability
distribution of network states under various conditions, such as under
spontaneous activity of neuronal networks \cite{shlens2006structure,tang2008maximum}
or visual input \cite{schneidman2006weak}. Since the second-order
MEP distribution can be obtained by maximizing the Shannon entropy
with measured constraints of only the first-order and the second-order
moments, the curse of dimensionality is circumvented and an accurate
estimation of the probability distribution of network states is provided.
A series of works have been initiated to address various aspects of
this approach, including explorations of fast algorithms \cite{nasser2013spatio,broderick2007faster},
the inference of spatial-temporal correlations \cite{tang2008maximum,shlens2008synchronized,marre2009prediction,yeh2010maximum,nasser2014parameter},
and the characterization of network functional connectivity \cite{yu2008small,roudi2009ising,hertz2011ising,watanabe2013pairwise,barton2013ising,dunn2015correlations}.

In addition to spiking neuronal networks \cite{schneidman2006weak},
the MEP analysis has been applied to analyzing various types of binary
data. These applications include, for example, functional magnetic
resonance imaging data \cite{watanabe2013pairwise}, in which each
brain region is considered to be either active or silent, and stock
market data \cite{bury2012statistical}, in which the price of each
stock is considered to be either increasing or not. However, for all
these binary data, long-time recordings are often impractical to be
obtained. 

In this work, we begin by showing that there exist invertible linear
transforms among the probability distribution of network states, the
moments, and the effective interactions in the MEP analysis for a
general network of nodes with binary dynamics. Based on these transforms,
we show that the second-order MEP analysis gives rise to an accurate
estimate of the probability distribution of network states with a
short-time recording. Finally, using data from both simulated HH neuronal
network model and the electrophysiological experiment, we demonstrate
the good performance of the second-order MEP analysis with short recordings.
Therefore, the applicability of the second-order MEP analysis in practical
situations could be significantly improved. 

Finally, we point out that there are also some limitations on low-order
MEP analysis. First, low-order MEP analysis is often insufficient
to accurately reconstruct the probability distribution of network
states in synchronized networks since the high-order effective interactions
in such cases are often not small \cite{xu2016dynamical}. Second,
as both the order of moment constraints and the network size, $n$,
increase, the existing algorithms to estimate effective interactions
for a large network become very slow \cite{shlens2009structure,nasser2013spatio}.
Because the number all network states, i.e., $2^{n}$, is too large
when $n$ is a large number, the existing algorithms estimate moments
of the MEP distribution using Monte Carlo sampling or its variants
from the MEP distribution, which are often very slow when the dimension
of the distribution is high \cite{shlens2009structure,nasser2013spatio}.
Therefore, how to efficiently characterize the statistical properties
of strongly-correlated, e.g., nearly synchronized, network dynamics
and how to develop fast numerical algorithms for the application of
the MEP analysis in large-scale networks remain interesting and challenging
issues for future studies.
\begin{acknowledgments}
This work was supported by NSFC-11671259, NSFC-11722107, NSFC-91630208
and Shanghai Rising-Star Program-15QA1402600 (D.Z.); by NSF DMS-1009575
and NSFC-31571071 (D.C.); by Shanghai 14JC1403800, 15JC1400104, and
SJTU-UM Collaborative Research Program (D.C. and D.Z.); and by the
NYU Abu Dhabi Institute G1301 (Z.X., D.Z., and D.C.).
\end{acknowledgments}

\appendix

\section{The iterative scaling algorithm \label{subsec:A2:-The-iterative}}

We briefly describe the widely-used numerical algorithm for the estimation
of effective interactions from moments. More details about this algorithm
can be found in Ref. \cite{tang2008maximum}. For illustration, we
present the procedure to obtain the second-order MEP distribution,
$P_{2}(\Omega)$. The interactions are initialized by: $J_{i}=\left\langle \sigma_{i}\right\rangle _{P(\Omega)}$
and $J_{ij}=\left\langle \sigma_{i}\sigma_{j}\right\rangle _{P(\Omega)}$,
where $\left\langle \cdot\right\rangle _{P(\Omega)}$ denotes the
expectation with respect to the measured probability distribution
of network states in data recording, $P(\Omega)$. The expected values
of the individual means $\left\langle \sigma_{i}\right\rangle _{P_{2}(\Omega)}$
and pairwise correlations $\left\langle \sigma_{i}\sigma_{j}\right\rangle _{P_{2}(\Omega)}$
with respect to the second-order MEP distribution $P_{2}(\Omega)$
can be determined by 
\[
\left\langle \sigma_{i}\right\rangle _{P_{2}(\Omega)}\equiv\sum_{l=1}^{2^{n}}\sigma_{i}(\Omega_{l})P_{2}(\Omega_{l}),
\]
\[
\left\langle \sigma_{i}\sigma_{j}\right\rangle _{P_{2}(\Omega)}\equiv\sum_{l=1}^{2^{n}}\sigma_{i}(\Omega_{l})\sigma_{j}(\Omega_{l})P_{2}(\Omega_{l}),
\]
where $\sigma_{i}(\Omega_{l})$ is the state of the $i$th node in
the network state $\Omega_{l}$. To improve the agreement between
$\left\langle \sigma_{i}\right\rangle _{P_{2}(\Omega)}$, $\left\langle \sigma_{i}\sigma_{j}\right\rangle _{P_{2}(\Omega)}$
and $\left\langle \sigma_{i}\right\rangle _{P(\Omega)}$, $\left\langle \sigma_{i}\sigma_{j}\right\rangle _{P(\Omega)}$,
the values of $J_{i}$ and $J_{ij}$ are adjusted by an iterative
procedure: 
\[
J_{i}^{new}=J_{i}^{old}+\alpha\mbox{sign}\left(\left\langle \sigma_{i}\right\rangle _{P(\Omega)}\right)\log\frac{\left\langle \sigma_{i}\right\rangle _{P(\Omega)}}{\left\langle \sigma_{i}\right\rangle _{P_{2}(\Omega)}},
\]
\[
J_{ij}^{new}=J_{ij}^{old}+\alpha\mbox{sign}\left(\left\langle \sigma_{i}\sigma_{j}\right\rangle _{P(\Omega)}\right)\log\frac{\left\langle \sigma_{i}\sigma_{j}\right\rangle _{P(\Omega)}}{\left\langle \sigma_{i}\sigma_{j}\right\rangle _{P_{2}(\Omega)}},
\]
where the constant $\alpha$ is used to maintain the stability of
the iteration. We use $\alpha=0.75$ as in Ref. \cite{tang2008maximum}.

\section{The Hodgkin-Huxley neuron model\label{sec:The-Hodgkin-Huxley-neuron}}

The HH model is described as follows. The dynamics of the membrane
potential of the $i$th neuron, $V_{i}$, is governed by \cite{sun2009library,sun2010pseudo}
\begin{align*}
C\frac{\mathrm{d}V_{i}}{\mathrm{d}t} & =I_{{\rm Na}}+I_{{\rm K}}+I_{{\rm L}}+I_{i}^{\textrm{input}},\\
I_{{\rm Na}} & =-(V_{i}-V_{\textrm{Na}})G_{\textrm{Na}}{\rm h}_{i}{\rm m}_{i}^{3},\\
I_{{\rm K}} & =-(V_{i}-V_{\textrm{K}})G_{\textrm{K}}{\rm n}_{i}^{4},\\
I_{{\rm L}} & =-(V_{i}-V_{\textrm{L}})G_{\textrm{L}},
\end{align*}
with 
\begin{equation}
\frac{\mathrm{d}X_{i}}{\mathrm{d}t}=(1-X_{i})\alpha_{X}(V_{i})-X_{i}\beta_{X}(V_{i}),\label{eq:HHdX}
\end{equation}
where the gating variable $X={\rm m},{\rm n},{\rm h}$ and 

\begin{center}
\begin{tabular}{ll}
${\displaystyle \alpha_{n}(V_{i})=\frac{0.1-0.01V_{i}}{\exp(1-0.1V_{i})-1}}$, & ${\displaystyle \beta_{n}(V_{i})=\frac{5}{40}\exp(-V_{i}/80)}$,\tabularnewline
${\displaystyle \alpha_{m}(V_{i})=\frac{2.5-0.1V_{i}}{\exp(2.5-0.1V_{i})-1}}$, & ${\displaystyle \beta_{m}(V_{i})=4\exp\left(-V_{i}/18\right)}$,\tabularnewline
${\displaystyle \alpha_{h}(V_{i})=0.07\exp(-V_{i}/20)}$, & ${\displaystyle \beta_{h}(V_{i})=\frac{1}{\exp(3-0.1V_{i})+1}}$.\tabularnewline
\end{tabular}
\par\end{center}

The current $I_{i}^{\textrm{input}}$ describes inputs to the $i$th
neuron coming from the external drive of the network, as well as interactions
between neurons in the network, $I_{i}^{\textrm{input}}=I_{i}^{\textrm{E}}+I_{i}^{\textrm{I}}$
with $I_{i}^{\textrm{E}}=-(V_{i}-V_{G}^{\textrm{E}})G_{i}^{\textrm{E}}$
and $I_{i}^{\textrm{I}}=-(V_{i}-V_{G}^{\textrm{I}})G_{i}^{\textrm{I}}$,
where $I_{i}^{\textrm{E}}$ and $I_{i}^{\textrm{I}}$ are excitatory
and inhibitory input currents, respectively, and $V_{G}^{\textrm{E}}$
and $V_{G}^{\textrm{I}}$ are their corresponding reversal potentials.
The dynamics of the conductance, $G_{i}^{Q}$, for $Q=\textrm{E},\textrm{I}$
are described as follows, 
\begin{align*}
\frac{\mathrm{d}G_{i}^{Q}}{\mathrm{d}t} & =-\frac{G_{i}^{Q}}{\sigma_{G}^{Q}}+H_{i}^{Q},\\
\frac{\mathrm{d}H_{i}^{Q}}{\mathrm{d}t} & =-\frac{H_{i}^{Q}}{\sigma_{H}^{Q}}+\sum_{k}F_{i}^{Q}\delta(t-T_{i,k}^{F})+\sum_{j\neq i}C_{ij}g(V_{j}^{\textrm{pre}}),
\end{align*}
with $g(V_{j}^{\textrm{pre}})=1/\left(1+\exp(-(V_{j}^{\textrm{pre}}-85)/2)\right),$
where $F_{i}^{Q}$ is the strength of the external \emph{Poisson input}
of rate $\mu_{i}$ to neuron $i$ with $T_{i,k}^{F}$ being the time
of the $k$th input event. We use $F_{i}^{\textrm{E}}=f$, $F_{i}^{\textrm{I}}=0$,
$\mu_{i}=\mu$ for all the neurons in our simulation. The parameter
$C_{ij}$ describes the coupling strength from the $j$th presynaptic
neuron to the $i$th neuron, and $V_{j}^{\textrm{pre}}$ is the membrane
potential of the $j$th presynaptic neuron. The adjacency matrix,
$\mathbf{W}=(w_{ij})$, describes the neuronal network connectivity
structure and $C_{ij}=w_{ij}C^{Q_{i}Q{}_{j}}$, where $Q_{i}$, $Q_{j}$
is chosen as $E$ or $I$, indicating the neuron type of the $i$th
neuron and the $j$th neuron ($C^{Q_{i}Q{}_{j}}$ is one of $C^{\textrm{EE}}$,
$C^{\textrm{EI}}$, $C^{\textrm{IE}}$, $C^{\textrm{II}}$). The value
$w_{ij}=1$ if there is a directed coupling from the $j$th presynaptic
neuron to the $i$th postsynaptic neuron and $w_{ij}=0$ otherwise. 

In this study, the values of parameters in the above conductance equations
are chosen as: $V_{\textrm{Na}}=\unit[115]{mV}$, $V_{\textrm{K}}=\unit[-12]{mV}$,
$V_{\textrm{L}}=\unit[10.6]{mV}$ (the resting potential of a neuron
is set to $\unit[0]{mV}$), $G_{\textrm{Na}}=\unit[120]{mS\cdot cm^{-2}}$,
$G_{\textrm{K}}=\unit[36]{mS\cdot cm^{-2}}$, $G_{\textrm{L}}=\unit[0.3]{mS\cdot cm^{-2}}$,
the membrane capacity $C=\unit[1]{\mu F\cdot cm^{-2}}$, $V_{G}^{\textrm{E}}=\unit[65]{mV}$,
$V_{G}^{\textrm{I}}=\unit[-15]{mV}$, $\sigma_{G}^{E}=\unit[0.5]{ms}$,
$\sigma_{H}^{E}=\unit[3.0]{ms}$, $\sigma_{G}^{I}=\unit[0.5]{ms}$,
and $\sigma_{H}^{I}=\unit[7.0]{ms}$. We keep the Poisson input parameters
fixed during each single simulation. 

In our numerical simulation, an explicit fourth order Runge-Kutta
method is used \cite{sun2009library,sun2010pseudo} with time step
$\unit[\sim0.03]{ms}$. The spike train data were obtained with a
sufficiently high sampling rate, e.g., $\unit[2]{kHz}$. 

\section{Proof of one-to-one mapping between the probability distribution
and moments \label{subsec:A3:-The-proof}}

We prove that there exists a full-rank matrix ${\bf U}_{{\rm PM}}^{(n)}$
that transforms from $\mathbf{P}^{(n)}$ to\emph{ $\mathbf{M}^{(n)}$,
}i.e., 
\begin{equation}
{\bf U}_{{\rm PM}}^{(n)}\mathbf{P}^{(n)}=\mathbf{M}^{(n)},\label{eq:AQR-1}
\end{equation}
 for a network of any size $n$. As an illustration with a network
of $n=2$ nodes, the expectation of $\sigma_{1}$, $\sigma_{2}$,
$\sigma_{1}\sigma_{2}$ can be obtained by 
\begin{equation}
\left[\begin{array}{cccc}
1 & 1 & 1 & 1\\
0 & 1 & 0 & 1\\
0 & 0 & 1 & 1\\
0 & 0 & 0 & 1
\end{array}\right]\left[\begin{array}{c}
P_{00}\\
P_{10}\\
P_{01}\\
P_{11}
\end{array}\right]=\left[\begin{array}{c}
1\\
\left\langle \sigma_{1}\right\rangle \\
\left\langle \sigma_{2}\right\rangle \\
\left\langle \sigma_{1}\sigma_{2}\right\rangle 
\end{array}\right],\label{eq:N2Cp-1}
\end{equation}
i.e., $\mathbf{U}_{{\rm PM}}^{(2)}\mathbf{P}^{(2)}=\mathbf{M}^{(2)}$.
Clearly, from Eq. (\ref{eq:N2Cp-1}), $\mathbf{U}_{{\rm PM}}^{(2)}$
is of full rank. We now prove the above result for any $n$ by mathematical
induction. Suppose $\mathbf{U}_{{\rm PM}}^{(k)}$ is of full rank.
Then $\mathbf{P}^{(k+1)}$ and $\mathbf{M}^{(k+1)}$ can be decomposed
into two parts with equal length of $2^{k}$ and $\mathbf{U}_{{\rm PM}}^{(k+1)}$
can be decomposed into four sub-matrices with the dimension of each
sub-matrix being $2^{k}\times2^{k}$ as follows: 
\[
\left[\begin{array}{cc}
\mathbf{U}_{11}^{(k+1)} & \mathbf{U}_{12}^{(k+1)}\\
\mathbf{U}_{21}^{(k+1)} & \mathbf{U}_{22}^{(k+1)}
\end{array}\right]\left[\begin{array}{c}
\mathbf{P}_{1}^{(k+1)}\\
\mathbf{P}_{2}^{(k+1)}
\end{array}\right]=\left[\begin{array}{c}
\mathbf{M}_{1}^{(k+1)}\\
\mathbf{M}_{2}^{(k+1)}
\end{array}\right],
\]
where 
\[
\mathbf{P}_{1}^{(k+1)}=\left[\begin{array}{c}
P_{000\cdots000}\\
P_{100\cdots000}\\
P_{010\cdots000}\\
P_{110\cdots000}\\
\vdots\\
P_{111\cdots110}
\end{array}\right],\quad\quad\mathbf{P}_{2}^{(k+1)}=\left[\begin{array}{c}
P_{000\cdots001}\\
P_{100\cdots001}\\
P_{010\cdots001}\\
P_{110\cdots001}\\
\vdots\\
P_{111\cdots111}
\end{array}\right].
\]
The expression of $\mathbf{M}_{1}^{(k+1)}$ is the same as the expression
of $\mathbf{M}^{(k)}$ in Eq. (\ref{eq:QR}), i.e., $\mathbf{M}_{1}^{(k+1)}=\mathbf{M}^{(k)}$,
and $\mathbf{M}_{2}^{(k+1)}$ can be expressed as follows, 
\[
\mathbf{M}_{2}^{(k+1)}=\left[\begin{array}{c}
\left\langle 1\cdot\sigma_{k+1}\right\rangle \\
\left\langle \sigma_{1}\sigma_{k+1}\right\rangle \\
\left\langle \sigma_{2}\sigma_{k+1}\right\rangle \\
\left\langle \sigma_{1}\sigma_{2}\sigma_{k+1}\right\rangle \\
\vdots\\
\left\langle \sigma_{2}\prod_{j=3}^{k}\sigma_{j}\sigma_{k+1}\right\rangle \\
\left\langle \sigma_{1}\sigma_{2}\prod_{j=3}^{k}\sigma_{j}\sigma_{k+1}\right\rangle 
\end{array}\right].
\]
In the representation of the base-2 number system of $k+1$ digits,
the first digit of the representation of $i-1$, indicating the state
of node $k+1$, is $0$ for $1\leq i\leq2^{k}$ and $1$ for $2^{k}+1\leq i\leq2^{k+1}.$
For any $i$ with $1\leq i\leq2^{k}$, suppose $p_{i}^{(k)}$ is the
probability of state $(\sigma_{1},\sigma_{2},\cdots,\sigma_{k})$,
i.e., $P_{\sigma_{1}\sigma_{2}\cdots,\sigma_{k}}$, then, the $i$th
entry of $\mathbf{P}_{1}^{(k+1)}$ and $\mathbf{P}_{2}^{(k+1)}$ are
$P_{\sigma_{1}\sigma_{2}\cdots,\sigma_{k},0}$ and $P_{\sigma_{1}\sigma_{2}\cdots,\sigma_{k},1}$,
respectively. Thus, the states of nodes from the first to the $2^{k}$th
are all the same for $\mathbf{P}^{(k)}$, $\mathbf{P}_{1}^{(k+1)}$
and $\mathbf{P}_{2}^{(k+1)}$. Since $\mathbf{M}_{1}^{(k+1)}$ is
the same as $\mathbf{M}^{(k)}$, i.e., their expressions only consider
nodes from the first to the $2^{k}$th, the contribution from both
$\mathbf{P}_{1}^{(k+1)}$ and $\mathbf{P}_{2}^{(k+1)}$ to $\mathbf{M}_{1}^{(k+1)}$
is the same as the contribution from $\mathbf{P}^{(k)}$ to $\mathbf{M}^{(k)}$,\emph{
}i.e., $\mathbf{U}_{11}^{(k+1)}=\mathbf{U}_{{\rm PM}}^{(k)}$ and
$\mathbf{U}_{12}^{(k+1)}=\mathbf{U}_{{\rm PM}}^{(k)}$. Since the
state of node $k+1$ is $0$ in all entries of $\mathbf{P}_{1}^{(k+1)}$,
there is no contribution of $\mathbf{P}_{1}^{(k+1)}$ to $\mathbf{M}_{2}^{(k+1)}$,
i.e., $\mathbf{U}_{21}^{(k+1)}=\mathbf{0}$.

Similarly, since the state of node $k+1$ is $1$ in all entries of
$\mathbf{P}_{2}^{(k+1)}$, the contribution from $\mathbf{P}_{2}^{(k+1)}$
to $\mathbf{M}_{2}^{(k+1)}$ only depends on the states of nodes from
the first to the $2^{k}$th---there is a one-to-one correspondence
of each entry between $\mathbf{P}_{2}^{(k+1)}$ and $\mathbf{P}^{(k)}$,
and between $\mathbf{M}_{2}^{(k+1)}$ and $\mathbf{M}^{(k)}$ as aforementioned.
Thus, $\mathbf{U}_{22}^{(k+1)}=\mathbf{U}_{{\rm PM}}^{(k)}$ and we
can obtain a \emph{recursive relation}, that is, 
\begin{equation}
\mathbf{U}_{{\rm PM}}^{(k+1)}=\left[\begin{array}{cc}
\mathbf{U}_{{\rm PM}}^{(k)} & \mathbf{U}_{{\rm PM}}^{(k)}\\
\mathbf{0}^{(k)} & \mathbf{U}_{{\rm PM}}^{(k)}
\end{array}\right],\label{eq:MomProRela-1}
\end{equation}
where $\mathbf{0}^{(k)}$ is the zero matrix with dimension $2^{k}\times2^{k}$.
Thus, $\mathbf{U}_{{\rm PM}}^{(k+1)}$ is also a matrix of full rank.
By induction, $\mathbf{U}_{{\rm PM}}^{(n)}$ is of full rank for any
$n$.

\section{Proof of the recursive relation among effective interactions \label{sec:Proof-of-Recursive}}

For a network of $n$ nodes, we prove the following recursive relation
in the full-order MEP analysis using mathematical induction:

\begin{equation}
J_{12\cdots(k+1)}=J_{12\cdots k}^{1}-J_{12\cdots k},\label{eq:Jobtain-1}
\end{equation}
where the term $J_{12\dots k}^{1}$ is obtained from the $k$th order
effective interaction $J_{12...k}$ by changing the state of the $(k+1)$st
node in it. 

From the mapping between effective interactions and the probability
distribution (Eq. (\ref{eq:BJQ})), it can be seen that the effective
interactions is simply a linear combination of logarithm of the probability
distribution. To express this explicitly, we first introduce the notation
$H_{m}^{l}$ as
\begin{equation}
H_{m}^{l}=\sum_{\Omega\in\Lambda_{m}^{l}}\log P(\Omega),\label{eq:Uml}
\end{equation}
where $\Lambda_{m}^{l}=\{(\sigma_{1},\sigma_{2},\cdots,\sigma_{n})|\sum_{i=1}^{m}\sigma_{i}=l;\sigma_{j}=0,m<j\leq n\}$,
$0\leq l\leq m\leq n$.

We \emph{then} show that if the $k$th-order ($1\leq k\leq n$) effective
interaction, $J_{12\cdots k}$, can be expressed as
\begin{equation}
J_{12\cdots k}=\sum_{i=0}^{k}(-1)^{k-i}H_{k}^{i},\label{eq:Juniform}
\end{equation}
Eq. (\ref{eq:Jobtain-1}) is also valid as follows. For $1\leq i\leq k$,
there are $i$ nodes out of $k+1$ nodes active in the states described
by $H_{k+1}^{i}$ of $\Lambda_{k+1}^{^{i}}$. For $0<i<k+1$, we can
split $H_{k+1}^{i}$ into two terms, one is $H_{k}^{i}$, where the
$(k+1)$st node is inactive, the other term is $W_{k}^{i-1}\equiv H_{k+1}^{i}-H_{k}^{i}$,
where the $(k+1)$st node is active. We also define $W_{k}^{k}\equiv H_{k+1}^{k+1}$.
From Eq. (\ref{eq:Juniform}), we have
\[
J_{12\cdots(k+1)}=\sum_{i=0}^{k+1}(-1)^{k+1-i}H_{k+1}^{i}.
\]
By using $H_{k+1}^{i}=W_{k}^{i-1}+H_{k}^{i}$ and $H_{k+1}^{0}=H_{k}^{0}$,
we have \begin{widetext}
\begin{align*}
J_{12\cdots(k+1)} & =H_{k+1}^{k+1}+\sum_{i=1}^{k}(-1)^{k+1-i}(W_{k}^{i-1}+H_{k}^{i})-(-1)^{k}H_{k+1}^{0}\\
 & =W_{k}^{k}+\sum_{i=1}^{k}(-1)^{k+1-i}W_{k}^{i-1}-(-1)^{k}H_{k+1}^{0}-\sum_{i=1}^{k}(-1)^{k-i}H_{k}^{i}\\
 & =\sum_{i=0}^{k}(-1)^{k-i}W_{k}^{i}-\sum_{i=0}^{k}(-1)^{k-i}H_{k}^{i}\\
 & =J_{12\cdots k}^{1}-J_{12\cdots k},
\end{align*}
\end{widetext}where $J_{12\cdots k}^{1}$ is the quantity which switches
the state of the $(k+1)$st node in the $k$th-order effective interaction
$J_{12...k}$ from inactive to active. Therefore, the recursive relation
(Eq. (\ref{eq:Jobtain-1})) is proved if Eq. (\ref{eq:Juniform})
is valid for $1\leq k\leq n$. 

We \emph{next} prove the validity of Eq. (\ref{eq:Juniform}) by mathematical
induction as follows. For $k=1$, as shown in Eq. (\ref{eq:hP}) in
the main text, we have 
\begin{align*}
J_{1} & =\log\frac{P_{10\cdots0}}{P_{00\cdots0}}\\
 & =H_{1}^{1}-H_{1}^{0}.
\end{align*}
Therefore, Eq. (\ref{eq:Juniform}) is valid when $k=1$. Now, we
assume Eq. (\ref{eq:Juniform}) is valid for $k\leq K$.

We next want to prove that Eq. (\ref{eq:Juniform}) holds for $k=K+1$.
We begin by show that an arbitrary $g$th-order ($g\leq K$) effective
interaction $J_{i_{1}\cdots i_{g}}$ can be expressed as follows:
\begin{equation}
J_{i_{1}\cdots i_{g}}=\sum_{i=0}^{g}(-1)^{g-i}H_{A_{g}}^{i},\quad g\leq K\label{eq:Juniform2}
\end{equation}
where $A_{g}=\{i_{1},\cdots,i_{g}\}$, 
\[
H_{A_{g}}^{l}=\sum_{\Omega\in\Lambda_{A_{g}}^{l}}\log P(\Omega),
\]
 and $\Lambda_{A_{g}}^{l}=\{(\sigma_{i_{1}},\sigma_{i_{2}},\cdots,\sigma_{i_{n}})|\sum_{i_{j}\in A_{g}}\sigma_{i_{j}}=l;\sigma_{i_{j}}=0,g<j\leq n\}$.
To show the validity of Eq. (\ref{eq:Juniform2}), we can permute
the neuronal indexes from $\{1,2,\cdots,n\}$ to $\{i_{1},i_{2},\cdots,i_{n}\}$
by the mapping $j\rightarrow i_{j}$ for $1\leq j\leq n$. Since Eq.
(\ref{eq:Juniform}) is valid for $g\leq K$ by assumption, Eq. (\ref{eq:Juniform2})
is also valid. 

Next, we study the relation between $J_{12\cdots(K+1)}$ and the effective
interactions whose orders are smaller than $K+1$. By substituting
$\Omega=(1,1,\cdots,1,0,\cdots,0)$ (nodes from $1$ to $K+1$ are
active and nodes from $K+2$ to $n$ are inactive) into the full-order
MEP analysis, we obtain
\begin{equation}
J_{12\cdots(K+1)}+\sum_{g=1}^{K}\sum_{i_{1}<\cdots<i_{g}}^{K+1}J_{i_{1}\cdots i_{g}}=H_{K+1}^{K+1}-H_{K+1}^{0}.\label{eq:Jmo}
\end{equation}
For $g\leq K$, from Eq (\ref{eq:Juniform2}), by the induction assumption,
we have
\begin{equation}
\sum_{i_{1}<\cdots<i_{g}}^{K+1}J_{i_{1}\cdots i_{g}}=\sum_{i=0}^{g}(-1)^{g-i}C_{K+1-i}^{g-i}H_{K+1}^{i},\label{eq:Jmid1}
\end{equation}
where $C_{K+1-i}^{g-i}$ is the number of the selection of $g-i$
terms from all the possible $K+1-i$ choices. Since $J_{i_{1}\cdots i_{g}}$
is the $g$th-order effective interaction, as in Eq. (\ref{eq:Juniform2}),
the sign of the logarithm probability of a state in which there are
$i$ nodes active is $(-1)^{g-i}$. To consider the coefficient of
$H_{K+1}^{i}$ in the right hand side of Eq. (\ref{eq:Jmid1}), we
can consider that for each group of $i$ nodes, how many groups of
$g$ nodes in the left hand side of Eq. (\ref{eq:Jmid1}) containing
these considered $i$ nodes, where $g\geq i$. If there are $g$ nodes
containing the considered $i$ nodes, then, there are only $g-i$
nodes unknown. These $g-i$ nodes can be chosen from the group of
nodes which belongs to the total $K+1$ nodes but not the considered
$i$ nodes. Therefore, the choice number of these $g-i$ nodes is
$C_{K+1-i}^{g-i}$. Then, we have
\begin{equation}
J_{12\cdots(K+1)}=H_{K+1}^{K+1}-H_{K+1}^{0}-\sum_{g=1}^{K}\sum_{i=0}^{g}(-1)^{g-i}C_{K+1-i}^{g-i}H_{K+1}^{i}.\label{eq:Jmo-1}
\end{equation}
For $K+1>l>0$, the coefficient of $H_{K+1}^{l}$ is 
\begin{equation}
-\sum_{g=l}^{K}(-1)^{g-l}C_{K+1-l}^{g-l}=(-1)^{K+1-l}.\label{eq:Jmo3}
\end{equation}
The coefficient of $H_{K+1}^{0}$ is
\begin{equation}
-1-\sum_{g=1}^{K}(-1)^{g}C_{K+1}^{g}=(-1)^{K+1}.\label{eq:Jmo2}
\end{equation}
Through Eqs. (\ref{eq:Jmo-1}), (\ref{eq:Jmo2}) and (\ref{eq:Jmo3}),
we obtain
\begin{equation}
J_{12\cdots(K+1)}=\sum_{i=0}^{K+1}(-1)^{K+1-i}H_{K+1}^{i}.\label{eq:JU2}
\end{equation}
That is, Eq. (\ref{eq:Juniform}) is valid for $k=K+1$. By induction,
we obtain that Eq. (\ref{eq:Juniform}) holds for any integer $k$
with $1\leq k\leq n$. Therefore, we prove the validity of Eq. (\ref{eq:Jobtain-1}).

\section{Electrophysiological experiments\label{sec:pvc8Exp}}

The data were collected in the Laboratory of Adam Kohn at the Albert
Einstein College of Medicine and downloaded from crcns.org (pvc-8)
\cite{Kohn2015Multi}. These data consist of multi-electrode recordings
from V1 in anesthetized macaque monkeys, while natural images and
gratings were flashed on the screen. Recordings were performed using
the \textquotedblleft Utah\textquotedblright{} electrode array and
spike-sorting algorithm was used to determine spike trains corresponding
to each single neuron. Natural images were presented at two sizes,
$3-6.7$ degrees and windowed to $1$ degree, to quantify surround
modulation. All stimuli ($956$ in total) were displayed in pseudo-random
order for $\unit[100]{ms}$ each, followed by a $\unit[200]{ms}$
uniform gray screen. Each stimulus was presented 20 times. Experimental
procedures and stimuli are fully described in Ref. \cite{coen2015flexible}. 

\bibliographystyle{SIAM}
\bibliography{MEPRE}

\end{document}